\documentclass[11pt]{article}
\usepackage{xr}
\usepackage[margin=1.2in]{geometry}
\RequirePackage[OT1]{fontenc}
\RequirePackage{amsthm,amsmath}
\usepackage[square,sort,comma,numbers]{natbib}
\RequirePackage[colorlinks,citecolor=blue,urlcolor=blue]{hyperref}
\RequirePackage{yk}
\usepackage{amsthm,amsmath,amssymb,enumerate,mathrsfs,float}
\usepackage{color,marg}
\usepackage{dsfont}
\usepackage[listings]{tcolorbox}
\tcbuselibrary{listings,theorems}
\usepackage{mathtools}
\usepackage[noend]{algpseudocode}
\usepackage{mathtools}
\usepackage{caption}
\usepackage{subcaption}
\usepackage{tikz}
\usetikzlibrary{shapes.geometric, arrows}
\usepackage[ruled,vlined]{algorithm2e}

\usepackage{graphicx} %Loading the package
\graphicspath{{Figures/}}
\tikzstyle{startstop} = [rectangle, rounded corners, minimum width=3cm, minimum height=1cm,text centered, draw=black, fill=red!30]

\tikzstyle{io} =[rectangle, rounded corners, minimum width=3cm, minimum height=1cm,text centered, draw=black, fill=blue!30]

\tikzstyle{process} = [rectangle, rounded corners, minimum width=3cm, minimum height=1cm,text centered, draw=black, fill=orange!30]
\tikzstyle{decision} = [rectangle, rounded corners, minimum width=3cm, minimum height=1cm,text centered, draw=black, fill=green!30]

\tikzstyle{arrow} = [thick,->,>=stealth]

\usepackage{hyperref}
\hypersetup{
    colorlinks=true,
    linkcolor=blue,
    filecolor=magenta,      
    urlcolor=cyan,
}%\usepackage{hypernat}

\renewcommand{\P}{{\mathbb P}}
\RequirePackage{xcolor}

\theoremstyle{definition}

\theoremstyle{remark}

\newtcbtheorem{mytheo}{Theorem}%
{colback=green!5,colframe=green!35!black,fonttitle=\bfseries}{th}

\newtcbtheorem{mylemma}{Lemma}%
{colback=blue!5,colframe=purple!35!black,fonttitle=\bfseries}{th}

\newtcbtheorem{myassmp}{Assumption}%
{colback=green!5,colframe=black!35!black,fonttitle=\bfseries}{th}

\newdimen\AAdi%
\newbox\AAbo%
\def\AAk#1#2{\setbox\AAbo=\hbox{#2}\AAdi=\wd\AAbo\kern#1\AAdi{}}%

\newcounter{rcnt}[section]

\def\argmin{\mathop{\rm argmin}}
\def\argmax{\mathop{\rm argmax}}

\usepackage{tikz}
\usetikzlibrary{arrows}

\title{Markovian And Non-Markovian Processes with Active Decision Making Strategies For Addressing The COVID-19 Pandemic}
\author{Hamid Eftekhari, Debarghya Mukherjee, \\
Moulinath Banerjee and Ya'acov Ritov}
\begin{document}
\maketitle
%\begin{frontmatter}
% \begin{center}
% \large
% {\bf PI: Moulinath Banerjee, Professor of Statistics}
% \end{center}

% \begin{center}
% \large
% {\bf Co PI: Ya'acov Ritov, Professor Statistics}
% \end{center}

%\runtitle{Spatio-temporal modelling of pandemics}

%\end{frontmatter}
\begin{abstract}
We study and predict the evolution of Covid-19 in six US states from the period May 1 through August 31 using a discrete compartment-based model and  prescribe active intervention policies, like lockdowns, on the basis of minimizing a loss function, within the broad framework of partially observed Markov decision processes. For each state, Covid-19 data for 40 days (starting from May 1 for two northern states and June 1 for four southern states) are analyzed to estimate the transition probabilities between compartments and other parameters associated with the evolution of the epidemic.These quantities are then used to predict the course of the epidemic in the given state for the next 50 days (test period) under various policy allocations, leading to different values of the loss function over the training horizon. The optimal policy allocation is the one corresponding to the smallest loss. Our analysis shows that none of the six states need lockdowns over the test period, though the `no lockdown' prescription is to be interpreted with caution: responsible mask use and social distancing of course need to be continued. The caveats involved in modeling epidemic propagation of this sort are discussed at length. A sketch of a non-Markovian formulation of Covid-19 propagation (and more general epidemic propagation) is presented as an attractive avenue for future research in this area. 
\end{abstract} 
\section{Introduction}
The Coronavirus COVID-19 pandemic is one of the gravest global public health crises that the world has seen in the past 100 years, with echoes of the Spanish Flu of the early 20'th century. A highly contagious respiratory virus, it triggers problematic immune responses in enough individuals, especially among those with certain co-morbidities and the elderly, while also ravaging the lungs and other vital organs. Governments have had to force extensive lockdowns all over the world in a bid to control its rampant spread, engendering in the process, a massive economic downturn, the long term effects of which are largely unpredictable as of even date. This combination of a medical and fiscal crisis has brought forth an unprecedented response from several sections of society, almost on a war-like footing, with medical professionals at the helm, along with epidemiologists, economists, and last but not least data scientists bringing their combined expertise together to grapple with this threat, e.g. \cite{sun2020understanding}, \cite{atkeson2020will}, \cite{li2020neuroinvasive}, \cite{li2020neuroinvasive}, \cite{mckibbin2020global}, \cite{song2020epidemiological} (and references therein) to name a few.

This paper aims to address the temporal evolution of epidemic diseases in a population, paying attention to realistic features of the COVID epidemic, in conjunction with an active [i.e. in real time] decision making strategy depending on the observables of the epidemic process. We model the flow of the epidemic using an SEIRD (Susceptible-Exposed-Infected-Recovered-Deceased) model, which is an extension of the traditional SIR model with two additional compartments, \emph{Exposed} and \emph{Deceased}, and the \emph{Infected} compartment is split into two sub-compartments: \emph{mildly infected} and \emph{severely infected}. Instead of analyzing a continuous time evolution via differential equations, we confine ourselves to a discrete time Markovian process, taking the point of view that days are a natural unit of measurement. We note that Markov processes have a long history in the study of infectious diseases, started by the work of Kermack and McKendrick \cite{kermack1927contribution, kermack1932contributions,kermack1933contributions} and more recent works \cite{angulo2012modeling, yaesoubi2011generalized, angulo2013spatiotemporal}.  Furthermore, dynamic policies for the control of Markov decision processes have also been studied in some detail \cite{yaesoubi2011dynamic},\cite{smallwood1973optimal}, \cite{sondik1978optimal}, \cite{gangwani2019learning}, \cite{chakraborty2013statistical}, \cite{krishnamurthy2016partially}.

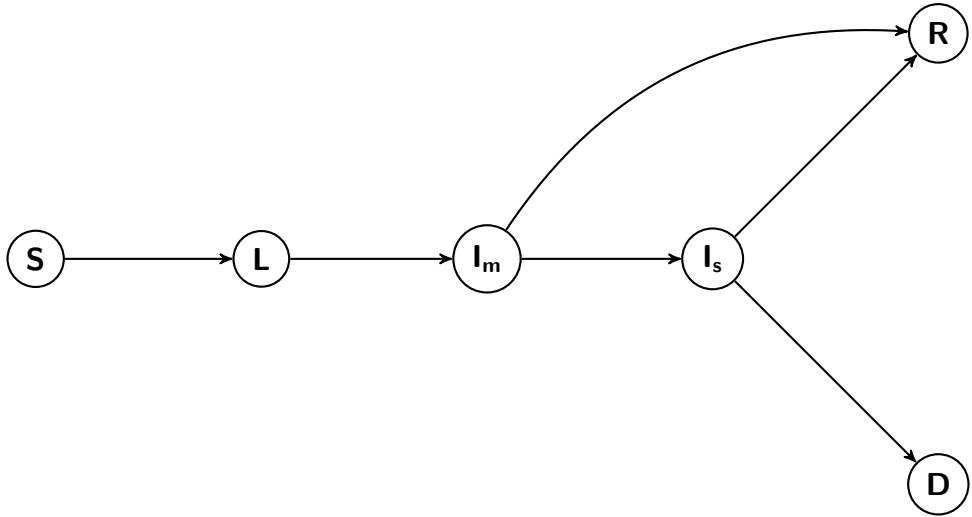
\begin{figure}
\centering
\begin{tikzpicture}[->,>=stealth',auto,node distance=3cm,
  thick,main node/.style={circle,draw,font=\sffamily\large\bfseries}]

  \node[main node] (1) {S};
  \node[main node] (2) [right of=1] {L};
  \node[main node] (3) [right of=2] {$\text{I}_\text{m}$};
  \node[main node] (4) [right of=3] {$\text{I}_\text{s}$};
  \node[main node] (5) [right of=4][above of=4] {R};
  \node[main node] (6) [right of=4][below of=4] {D};

  \path[every node/.style={font=\sffamily\large}]
    (1) edge node [right] {} (2)
    (2) edge node [right] {} (3)
    (3) edge node [right] {} (4)
    (3) edge[bend left] node [right] {} (5)
    (4) edge node [right] {} (5)
    (4) edge node [right] {} (6);
%    (3) edge[bend right] node [left] {} (1)
%    (4) edge[bend left] node [left] {} (1);
\end{tikzpicture}
\caption{\textcolor{blue}{Flow of the Markovian process: $S \leftarrow$ Susceptible, $L \leftarrow$ Latent/Exposed, $I_m \leftarrow$ Mildly infected, $I_s \leftarrow $Severely infected, $D \leftarrow$ Deceased, $R \leftarrow$ Recovered. An individual can move from compartment $A$ to compartment $B$ if $A \to B$.}}
\label{img:Markov_Flow}
\end{figure}

We say that a compartment $B$ is accessible from another compartment $A$ (write $A \to B$) if it is possible to move from $A$ to $B$ directly without hitting other compartments. For example, if we denote by $S$ the compartment of susceptible people and by $E$ the compartment of latent/exposed people then it is immediate that $S \to E$, but $S \nrightarrow R$ where $R$ is the compartment of recovered people, as a person cannot go to the recovery compartment directly from the susceptible one without hitting $E$: see figure \ref{img:Markov_Flow} for a detailed diagram of flow. We assume that at each time point, an individual has two options: either stay in the same compartment as they were in at the previous time, or move to another reachable compartment. The probability of transition to the next state is universal in the sense that it does not depend on the individuals but may depend on 
the previous state. This mechanism generates a Markov process on the number of individuals in each compartment. Some core features of our proposed Markovian model to be elaborated on in Section \ref{sec:Markov}  are: 
\begin{itemize}
\item[(a)] Allowing distinction between mild and severe infections and the subsequent evolution of such individuals in time.  
\item[(b)]  Explicit incorporation of an \emph{observed process} generated from a \emph{partially observed parent process} via a simple testing based mechanism. 
\item[(c)]  Active decision-making strategies based on the history of the epidemic. 
\item[(d)] Optimizing decision making strategies in terms of an expected loss/reward function that measures the trade-off between the economic costs of  lockdowns and the cost of lost lives or costs due to treatment denial. Clarification and detailed analysis of each of these aspects are provided in Section \ref{sec:Markov}. 
\end{itemize} 
We also present some ideas about a non-Markovian process that models each individual separately allowing their personal characteristics to determine their evolution, as a potential recipe for future studies. The non-Markovian model captures spatial heterogeneity by allowing the transition probability from susceptible to latent to depend on the distance between individuals. Section \ref{sec:Non_Markovian} deals with such details. Section \ref{sec:extension} concludes with a discussion of various possible directions for future research. 
\section{Markovian Model}
\label{sec:Markov}
Our first, and more parsimonious approach, is related to the paradigm of Partially Observed Markov Decision Processes (henceforth POMDP) which have been used extensively in monitoring processes evolving over a state space with active intervention and decision making strategies (see \cite{smallwood1973optimal}, \cite{sondik1978optimal}, \cite{gangwani2019learning}, \cite{chakraborty2013statistical}, \cite{krishnamurthy2016partially} and references therein). %We first present a more basic version of our model that captures key features of the COVID evolution dynamics, showcase some of our findings, and then discuss how the model will be embellished and extended to incorporate other realistic features.

%\begin{center}
%\begin{tikzpicture}[scale=0.5,->,>=stealth',auto,node distance=3cm,
%  thick,main node/.style={transform shape}]
%
%  \node[main node] (1) {S};
%  \node[main node] (2) [right of=1] {L};
%  \node[main node] (3) [right of=2] {$I_m$};
%  \node[main node] (4) [right of=3] {$I_s$};
%  \node[main node] (5) [right of=4][above of=4] {R};
%  \node[main node] (6) [right of=4][below of=4] {D};
%
%  \path[every node/.style={font=\sffamily\small}]
%    (1) edge node [right] {} (2)
%    (2) edge node [right] {} (3)
%    (3) edge node [right] {} (4)
%    (3) edge[bend left] node [right] {} (5)
%    (4) edge node [right] {} (5)
%    (4) edge node [right] {} (6);
%%    (3) edge[bend right] node [left] {} (1)
%%    (4) edge[bend left] node [left] {} (1);
%\end{tikzpicture}
%\end{center}
\subsection{Model structure }
Consider the following model for an epidemic: $S$ denotes the class of susceptibles, $L$ denotes latent individuals who are infected but cannot infect others as yet, $I_m$ comprises the (mildly) infected, $I_s$ the severely ill, $R$ those that have recovered and $D$ the deceased from the disease. The transition chain is from $S$ via $L$ to $I_m$ and from $I_m$ either to $R$ or to $I_s$ and from $I_s$ either to $R$ or $D$. A susceptible person only gets infected upon contact with an infected person, but not through a latent individual. Define $N$ as the total number of people in a fixed location which will be assumed to stay constant over the time horizon. The state space for this Markov process is $\mathcal{X} = \mathbb{N}^{6}$, where a generic element of $\cX$ should be thought as a vector of numbers of individuals in each compartment. The true numbers in the different compartments are driven by an underlying (\emph{hidden}, henceforth to be referred to as \emph{population}) Markov process which is not observed directly . We only observe (random or deterministic) fractions of these numbers, which gives rise to the \emph{observed} Markov process. Write the population Markov process as  $\{X_t \in \bbN^6\}_{t =1, \dots, T}$ over a finite horizon of time, where $X_{t, j}$ denotes the number of individuals in compartment $j \in \{S, L, I_m, I_s, R, D\}$ at time $t$. We denote the observed Markov Process as $X_t^O$ described next: On each day, we observe a cohort of mildly infected people and a cohort of severely ill individuals via testing. The former are typically asked to quarantine at home and self-treat, while the latter require immediate medical assistance and are hospitalized subject to capacity constraints. Furthermore, we record a fraction of recovered and deceased people, who are documented. The number of individuals that showed up for testing and tested negative [and can be taken to belong to the $S \cup L$ group] are not tracked. We assume therefore that we have a partially observed process: $X_t^O \in \bbN^4$ where $X^O_t = \{X^O_{t, j}\}_{j \in I_m, I_s, R, D}$. We take our action space to be $\mathcal{A} = \{0, 1, 2\}$ where $0$ indicates no lockdown, $1$ indicates partial lockdown, i.e. some essential businesses and organizations stay open and $2$ denotes a full lockdown, where only the most essential activities are sustained. 

Before proceeding further, let us articulate our convention about compartment numbers: any compartment number at a given time $t$ is taken to be the total count recorded by the end of day $t$, thus the number in a compartment at (the end of) day $t+1$ is given by the number at day $t$ plus movement into and out of the compartment in day $t+1$ where the parameters for such movements are either time invariant or depend upon the numbers recorded (at the end of) day $t$. The action $a_t$ (which is 0, 1 or 2) is proposed at the end of day $t$ and affects the infection dynamics from day $t+1$ onwards: $a_t = 0$ corresponds to `no lockdown' which we interpret broadly to mean that most activities are in place though people are practicing generally responsible behavior by and large, while $a_t = 1$ means `mild lockdown' where a substantial number of activities are closed, and $a_t = 2$ means `severe lockdown' where most activities have been restricted and human contact is minimal . Policy (decision) making, i.e. which action to take at what time, is based on the numbers for the \emph{observed process} via an optimization scheme based on a reward. 

%on Furthermore, we want to mention that, our model also allows finer clustering among the population of each space unit based on their medical condition. \DM{How much should we elaborate on that
\subsection{Population process and transition probabilities }
\label{sec:process}

The transition probability from the Susceptible to the Latent state on day $t+1$ is given by
$$P_{a_t,t+1}^{S \to L} = 1 - \exp\left(-R_0^{a_t} (P^{I_m \to I_s} + P^{I_m \to R})\frac{X_t(I_m)}{N} \right)\,.$$
%\begin{align*}
%P^{S \to L}_{(t+1)} = 1 - \exp\left( - \frac{R^{a_t} \cdot X_{t, S} \cdot X_{t, I_m} }{ (P_t^{I_m \to I_s} + P_t^{I_m \to R}) \cdot N^2 } \right),
%\end{align*}
where 
\begin{align*}
R_0^{a_t} = \begin{cases}
R_0 & \text{ if }  a_t = 2\\
R_0 + r_1 & \text{ if }  a_t = 1\\
R_0 + r_2 & \text{ if }  a_t = 0,
\end{cases}
\end{align*}
with $R_0$ denoting the basic reproductive number under severe lockdown and $r_1 <  r_2$. The basic reproduction number during any fixed lockdown phase is the average number of persons that an infected person transmits in that phase over the course of their illness when the proportion of susceptibles is near 1, i.e. the epidemic is in a controlled phase. Even though the early estimates of the basic reproductive number (under no lockdown) were between 2.43 to 3.10 \cite{d2020assessment}, we assume that over our period of analysis [May 2020 onwards by which time the epidemic had spread considerably at least in the northern US states] the value of $R_{a_t}$ remains smaller than $2$ due to increased awareness in society. Specifically, we assume that after reopening, the value of $R_{a_t}$ is 1.8 under no lockdown, while it assumes the values 1.3 and 0.8 respectively (drops of 0.5) under the first and second levels of lockdown.\footnote{Estimating the effect of lockdowns on the basic reproductive number ($r_1$ and $r_2$) is difficult given the available data, so we set $r_1 = 0.5$ and $r_2 = 1$ for our analysis. In principle, one can also estimate these parameters if more data is available.}The probabilties $P^{I_m \to I_s}$ and $P^{I_m \to R}$ are the chance of transiting from the $I_m$ state to states $I_s$ and $R$ respectively on day $t$ and are not taken to depend on $t$. The mathematical derivation of this formula is given in Section 5. 
%The intuition behind the displayed expression for the transition probability is as follows: We assume a person is contagious largely only when they are in $I_m$, but not in the latent/incubation stage or when they are in $I_s$ where they are typically immobile and under care. Note that the expected number of days a person remains in state $I_m$ is $1/(P_t^{I_m \to I_s} + P_t^{I_m \to R})$. The probability of infecting others will be proportional to the product of the proportion of susceptible people at time $t, X_{t, s}/N$, proportion of mildly infected people at time $t, X_{t, I_m}/N$, the basic reproduction number $R^{a_t}$ i.e. how many people a person can infect and on average how many days an infected person remains contagious i.e. $1/(P_t^{I_m \to I_s} + P_t^{I_m \to R})$. This gives rise to the above equation and the mathematical details are provided in the supplement. 

For the other transition probabilities, we assume that on any given day, an individual moves from $L$ to $I_m$ with probability $P^{L \to I_m}$. Similarly an individual in $I_m$ will either move to $I_s$ (w.p. $P^{I_m \to I_s}$), or to $R$ (w.p. $P^{I_m \to R}$) or stay in $I_m$ (w.p. $1 - P^{I_m \to I_s} - P^{I_m \to R}$). For an individual in $I_s$, they can go to $R$ (w.p. $P^{I_s \to R}$), or to $D$ (w.p. $P^{I_s \to D}$) or stay in $I_s$ (w.p. $1 - P^{I_s \to R} - P_t^{I_s \to D})$. The transition probability from $I_s$ to $D$ depends on time $t$ through the number of severely ill people in that location. If this exceeds the medical capacity, a severely ill person cannot be accommodated and the subsequent lack of medical care will affect the prognosis of illness, leading to a higher probability of death. Specifically, we use a simple dichotomous form: 
$$\textcolor{blue}{P^{I_s \to D}_{t+1} = P^{I_s \to D}_{1}\mathds{1}(X_{t, I_s} \le \mbox{cap}) + P^{I_s \to D}_{2}\mathds{1}(X_{t, I_s} >  \mbox{cap})} \,,$$
where $\mbox{cap}$ is hospitalization capacity at the location, and $P^{I_s \to D}_{2}$ is generally taken to be a multiple of $ P^{I_s \to D}_{1}$ (which we took to be 3 in our implementation) for our presented analysis. The other two compartments $R$ and $D$ are \emph{terminal} compartments, i.e. if a person is in one of these compartments, their status changes no further. The transition process from $X_{t}$ to $X_{t+1}$ under action $a_t$ can be summarized via the following mechanism: 
\begin{enumerate}
    \item Generate the random variables $Y_1, Y_2, Y_3, Y_4, Y_5, Y_6$ as (henceforth Bin and Mult are abbreviations of Binomial and Multinomial respectively):
\textcolor{blue}{
\begin{align*}
Y_1 & \sim Bin\left(X_{t, S}, P^{S \to L}_{a_t, t+1}\right) \\
Y_2 & \sim Bin\left(X_{t, L}, P^{L \to I_m}\right) \\
(Y_3, Y_4) & \sim Mult\left(X_{t, I_m}, P^{I_m \to I_s}, P^{I_m \to R}\right) \\
(Y_5, Y_6) & \sim Mult\left(X_{t, I_s}, P^{I_s \to R}, P^{I_s \to D}_{t+1}\right)
\end{align*}
}
\item Update the state: 
\textcolor{blue}
{
\begin{align*}
X_{t+1, S} & = X_{t, S} - Y_1, \\
X_{t+1, L} & = X_{t, L} + Y_1 - Y_2, \\ X_{t+1, I_m} & = X_{t, I_m} + Y_2 - Y_3 - Y_4, \\
X_{t+1, I_s} & = X_{t, I_s} +Y_3 - Y_5 - Y_6,\;\; \\
X_{t+1, R} & = X_{t, R} + Y_4 + Y_5,\\ X_{t+1, D} & = X_{t, D} + Y_6.
\end{align*}
}
\end{enumerate}

\subsection{Observed process and parameter estimation }
The observed process denoted by $X_t^O$ is driven by two mechanisms: First, the observed compartments themselves follow the same transitions as the population process, and second, at each point in time there is a flow from the population process to the observed: a number of individuals are tested and added to the observed compartments. Note that an individual in $X_{t, I_m}$ but not in $X_{t, I_m}^0$ (respectively in $X_{t, I_s}$ but not in $X_{t, I_s}^0$) will enter into $X^O_{t+1, I_m}$ (respectively into $X^O_{t+1, I_s}$) if they are tested positive. We assume that the number of individuals that test positive on day $t+1$ and are mildly infected (e.g. do not require immediate medical attention) follows a binomial distribution with parameters $X_{t, I_m} - X^O_{t, I_m}$ and $P_{t+1}(\text{presents for testing}\mid \text{mildly infected})$, henceforth denoted by $P^{test, mild}_{t+1}$. Similarly the number of individuals that test positive on day $t+1$ and are severely infected (e.g. need immediate medical attention) follows a binomial distribution with parameters $X_{t, I_s} - X^O_{t, I_s}$ and $P_{t+1}(\text{presents for testing}\mid \text{severely infected})$, henceforth denoted by $P^{test, severe}_{t+1}$. The process evolution can be summarized as follows:
\begin{enumerate}
    \item Generate the following random variables: 
    \textcolor{blue}{
    \begin{align*}
(Z_{I_m \to I_s} , Z_{I_m \to R} ) &\sim Mult\left( X^{O}_{t, I_m}, P^{I_m \to I_s}, P^{I_m \to R}\right) \\
(Z_{I_s \to R}, Z_{I_s \to D} ) &\sim Mult\left( X_{t, I_s}^{O}, P^{I_s \to R}, P_{t+1}^{I_s \to D}\right) \\
T_m(t+1) & \sim Bin\left(X_{t, I_m} - X_{t, I_m}^{O}, P^{test, mild}_{t+1}\right)\\
T_s(t+1) & \sim Bin\left(X_{t, I_s} - X_{t, I_s}^{O}, P^{test, severe}_{t+1}\right) 
\end{align*}
    }
    \item Then update the observed process as: 
    \textcolor{blue}{\begin{align*}
X_{t+1, I_m}^{O} &= X^{O}_{t, I_m} - Z_{I_m \to I_s} - Z_{I_m \to R} + T_m(t+1),\\
X_{t+1, I_s}^{O} &= X_{t, I_s}^{O} - Z_{I_s \to R} - Z_{I_s \to D} + Z_{I_m \to I_s} + T_s(t+1) \\
X_{t+1, R}^{O} &= X_{t, R}^{O} + Z_{I_m \to R} + Z_{I_s \to R}\\
X_{t+1, D}^{O} &= X_{t, D}^{O}  + Z_{I_s \to D} \,.
\end{align*}
}
\end{enumerate}
In order to simulate such a process over a time horizon $t = 1, \dots, T$, we need four sets of quantities: 
\begin{enumerate}
\item The numbers in the compartments of the population process at day 1. 
\item The numbers in the compartments of the observed process at day 1. 
\item Vector of transition probabilities between the compartments. 
\item Testing probabilities i.e. $P^{test, mild}_t, P^{test, severe}_t$. 
\end{enumerate}
We will discuss points 1 and 2 later, but briefly speaking, we set them based on real data and some inflation factor (viewing the population process as an inflated version of the observed process). So, assume for the moment that we have $X_1$ and $X^O_1$. Coming to points 3 and 4, we estimate the parameters (both transition probabilities and testing probabilities) jointly, based on the real data. To sketch the idea, we essentially use a co-ordinate descent approach, i.e. we update the transition probability vector and testing probabilities iteratively. Given an initial vector of transition probabilities, $P_0$,  one can generate the population process over the training period as described in Section 2.2. Now, as far as the observed process is concerned, note that we cannot generate $T_m, T_s$ as we don't know the testing probabilities yet. An intuitive and effective way is to replace them by \emph{the real data on daily new cases}, appropriately divided into mild and severe compartments: say at day $t$, we see $\hat T_m = $ new mild cases and $\hat T_s = $ new severe cases. Starting from Day 1, and using $\{(\hat T_m(t), \hat T_s(t))\}$ we can generate a version of the observed process, say $\hat X_t^O$ over the training period via steps (1) and (2) displayed above in this section [obviously ignoring the last two lines of (1)]. Next, the testing probabilities are estimated via the maximum likelihood principle as the ratio of observed mild cases and observed severe cases to the number of unobserved mild and severe cases respectively, since the denominators can be recovered from the infected numbers for the hidden and observed processes. 
%$$\hat P_t^{tested, mild} = \frac{\hat T_m(t)}{(X_{t, I_m} - \hat X_{t, I_m}^{O})}\;\;\mbox{and}\;\; \hat P_t^{tested, severe} = \frac{\hat T_s(t)}{(X_{t, I_s} - \hat X_{t, I_s}^{O})}$$ which are the maximal likelihood estimators of the success parameter for a binomial distribution.
 
Given these testing probabilities one can simulate the observed process $X_t^0$ for any transition probability vector $P$ via steps (1) and (2) above, and determine the one which emulates the real process the best, i.e. minimizes a loss function measuring the discrepancy between $X_t^0$ and the real data process. We now set the value of this minimizer to $P^0$ and repeat the two updating steps. We keep on updating both parameters over a fixed number of iterations and choose our final estimate as the one corresponding to that iteration which gives the smallest minimum loss over all iterations. The following algorithm summarizes this discussion:  

\begin{algorithm}[H]
\SetAlgoLined
 \textbf{Input: } Initial estimate of transition probabilities $\hat{P}_\text{init}$ and the number of iterations $K$ \;
Set $\hat{P}^{(0)} = \hat{P}_\text{init}$\;

\For {$k = 1, \dots, K$}{
   Given $\hat{P}^{(k-1)}$, obtain estimate of test probabilities $P^{test, mild(k)}_t$ and $P^{test, severe(k)}_t$ using Algorithm \ref{alg:tests}.\\
   \vspace{0.1in}
   Using $P^{test, mild(k)}_t$ and $P^{test, severe(k)}_t$ , obtain the updated transition probabilities $\hat{P}^{(k)}$ by minimizing the loss in Equation (\ref{eq:ls_criterion}) and let $\text{loss}^{(k)}$  be the value of this minimized loss.
   \vspace{0.1in}
 }
 Set $k_{\min} = \argmin_k \text{loss}^{(k)}$.\\
 Return $\hat P \equiv \hat{P}^{(k_{\min})}$ and $P^{test, mild(k_{\min})}_t$ and $P^{test, severe(k_{\min})}_t$.
 \caption{Joint Estimation of Transition and Test Probabilities}\label{alg:joint}
\end{algorithm}

We next describe each part of the iterative update procedure in detail. We start with the method of obtaining estimates of $P^{test, mild}_t$ and $P^{severe, mild}_t$ given a set of values $P$ for the transition probabilities.  Given $T(t)$ new cases on day $t$ based on real data \cite{jhu2020data}, we set 
%set We estimate these probabilities by generating values for both the hidden and observed processes over time [using initial estimates of transition probabilities to start the process] informed by the real data \cite{jhu2020data} on daily new cases, 
\[ \hat T_m(t)  = p_m T \;\; \mbox{and} \;\;\hat T_s(t) = p_s T \] 
where $p_s$ is the proportion of severe cases and $p_m \equiv 1 - p_s$ the proportion of mild cases. Exact values for these proportions are presented in Section \ref{section:statewise} where we do a state by state analysis. The generation of the process $\hat X_t^0$ incorporating real data is described next: 

\begin{enumerate}
\item Generate the random variables $Y^O_1, Y^O_2, Y^O_3, Y^O_4$ as:
\textcolor{blue}{
\begin{align*}
(Y^O_1, Y^O_2) & \sim Mult\left(X^O_{t, I_m}, P^{I_m \to I_s}, P^{I_m \to R}\right) \\
(Y^O_3, Y^O_4) & \sim Mult\left(X_{t, I_s}, P^{I_s \to R}, P^{I_s \to D}_{t}\right)
\end{align*}
}
\item Update the state: 
\textcolor{blue}
{
\begin{align*}
\hat X^0_{t+1, I_m} & = \hat X^O_{t, I_m} - Y^O_1 - Y^O_2 + \hat T_m(t+1), \\
\hat X^O_{t+1, I_s} & = \hat X^O_{t, I_s} + Y^O_1 - Y^O_3 - Y^O_4 + \hat T_s(t+1), \\
\hat X^O_{t+1, R} & = \hat X^O_{t, R} +Y^O_2 + Y^O_3,\;\; \\
\hat X^O_{t+1, D} & = \hat X^O_{t, D} + Y^O_4.
\end{align*}
}

\end{enumerate}

The processes $\{X_t\}$ and $\{\hat X^O_t\}$ are simulated for $T$ days where $T$ is the time length of the training period [taken to be 40 days] and the testing probabilites calculated as the proportion of daily new cases to unobserved cases, as in the below algorithm.  
%as it is MLE under binomial assumption. For example we can determine $p^{test, mild} = (\text{new mild cases}(t)/(X_{t-1}(I_m) - \hat X^O_{t-1}(I_m)))
%This is summarized in the following algorithm.      
%Then we run the process (both hidden and observed) and look at the difference between $X_{t-1, I_m} - X^O_{t-1, I_m}$, which quantifies the number of unobserved mild infection cases. Now one can estimate $P^{test, mild}$ by looking at the ratio of new mild cases to unobserved mild cases. Estimate of $P^{test, severe}$ can also be obtained in a similar fashion. The algorithm can be briefly described as follows: 

\begin{algorithm}[H]
\SetAlgoLined
 \textbf{Input: } Start with the data on daily new cases for $T$ consecutive days and estimates of transition probabilities between compartments\;
Set $\hat T_m(t) = T(t)  * p_m$ \;
Set $\hat T_s(t) = T(t) * p_s$ \;
Generate the processes $X_t$ and $\hat X^O_t$ for days 2 through $T$ with initial values for $X_1$ and $\hat X_1^O$\;
%Set t = 1 \; 
\While{$2 \leq t \leq T$}{
   Set $P^{test, mild}_{t} = \frac{\hat T_m(t)}{X_{t-1, I_m} - \hat X^O_{t-1, I_m}}$\,\;
   \vspace{0.1in}
   Set $P^{test, severe}_{t} = \frac{\hat T_s(t)}{X_{t-1, I_s} - \hat X^O_{t-1, I_s}}$\,\;
   \vspace{0.1in}
   Update $t \leftarrow t+1$. 
 }
 \caption{Calculating testing probabilities}\label{alg:tests}
\end{algorithm}
The initialization of the two processes as well as the elicitation of the initial transition probabilities will be discussed later. Observe that the above algorithm gives us testing probabilities for days $2$ through $T$. Note that, testing probabilities on day 1 are not required as we are not generating data on day 1, we are fixing it based on real data.

We next estimate the parameters $P = (P^{L\to I_m}, P^{I_m\to R}, P^{I_m\to I_s}, P^{I_s \to R}, P_1^{I_s \to D})$ (the inter-compartment transition probabilites) of the epidemic propagation\footnote{Note that the probability of transition from $S$ to $L$ at any given time is known once these probabilities and the action status $a$ are known.}. This is based on a discrepancy calculation: For days 1 through $T$, we have real data available on the number of infected as well as the number of epidemic related deaths in the location of interest. Note that the first is not a cumulative quantity, while the second is. This provides our \emph{real observed data process}: $(X_{t, active}^{real}, X_{t,D}^{real})$ for $1 \leq t \leq T$. On the other hand, for any parameter vector $P$, we can generate the population process $X_t$ from 1 through $T$, and also the observed process $X^{O}_t$ using the testing probabilities just determined. 
\begin{comment}
\begin{enumerate}
    \item Generate the following random variables: 
    \textcolor{blue}{
    \begin{align*}
(Z_{I_m \to I_s} , Z_{I_m \to R} ) &\sim Mult\left( X^{O}_{t, I_m}, P^{I_m \to I_s}, P^{I_m \to R}\right) \\
(Z_{I_s \to R}, Z_{I_s \to D} ) &\sim Mult\left( X_{t, I_s}^{O}, P^{I_s \to R}, P^{I_s \to R}\right) \\
T_m  & \sim Bin\left(X_{t, I_m} - X_{t, I_m}^{O,SIm}, P^{test, mild}_t\right)\\
T_s & \sim Bin\left(X_{t, I_s} - X_{t, I_s}^{O,Sim}, P^{test, severe}_t\right) 
\end{align*}
    }
    \item Then update the observed process as: 
    \textcolor{blue}{\begin{align*}
X_{t+1, I_m}^{O} &= X^{O}_{t, I_m} - Z_{I_m \to s} - Z_{I_m \to R} + T_m,\\
X_{t+1, I_s}^{O} &= X_{t, I_s}^{O} - Z_{I_s \to R} - Z_{I_s \to D} + Z_{I_m \to s} + T_s \\
X_{t+1, R}^{O} &= X_{t, R}^{O} + Z_{I_m \to R} + Z_{I_s \to R}\\
X_{t+1, D}^{O} &= X_{t, D}^{O}  + Z_{I_s \to D} \,.
\end{align*}
}
\end{enumerate}
Note that the random variables $Z_{I_m \to I_s}, Z_{I_m \to R}, Z_{I_s \to R}, Z_{I_s \to D}$ represents the daily transitions between observed compartments, whereas $T_m, T_s$ denotes the imported new cases from the unobserved hidden process via testing. 
\end{comment} 

The process $X^{O}$ should be differentiated from the process $\hat X^{O}$ from the previous section, as there is no real data directly involved in the former's construction. The goal of generating this new process is to obtain the infected and deceased numbers at each time $t$ and then compare them to the corresponding numbers for the real observed process. The parameter vector that minimizes the average discrepancy between 
$$ \left \{(X_{t, active}^{real}, X_{t,D}^{real}) \right\}_{t=1}^T \;\; \mbox{and} \;\; \left \{(X^{O}_{t, active}, X^{O}_{t,D}) \right\}_{t=1}^T $$  
is our estimate.\footnote{We deliberately leave recovery compartment from the parameter estimation procedure due to lack of real data on recovery for some of the states that we analyzed. With more data one can incorporate this in the procedure to obtain a better fit.}
%Denote the real observed data process by $X^{real}$ and the simulated observed process by $X^O$ as before. $X^{real}$ includes the deaths, recovered and active cases comparments for $T$ days. $X^{real}_{t,D}$ and $X^{real}_{t,R}$ are cumulative numbers whereas $
Note that $X^{O}_{t, active} = X^{O}_{t, I_m} + X^{O}_{t, I_s}$. 
\newline
\newline
More specifically, we first obtain the smoothed daily numbers of deaths as follows. Since $X_{t,D}^{O}$ is the cumulative number of deaths up to time $t$,  the 7-day moving average of daily deaths is obtained via
\begin{align*}
    D^{smooth\, obs.}_t = \frac{1}{7} \sum_{s = t - 3}^{t+3} X_{s, D}^{O} - X_{s-1, D}^{O} = \frac{ X_{t+3, D}^{O} - X_{t - 4, D}^{O}}{7}.
\end{align*}
Similarly, the smoothed daily number of deaths is obtained via $D^{\text{smoothed real}}_t = (X_{t+3, D}^{real} - X_{t - 4, D}^{real})/7$.

The following loss is now optimized over a grid of parameter values:
\begin{align}\label{eq:ls_criterion}
L(R_0, P, \varphi) = \Ex \left[ \sum_{t=1}^{T} \left( \frac{D_t^{\text{smooth obs}}}{D_t^{\text{smooth real}}}  - 1\right)^2
+ \left( \frac{ X_{t,I_m}^O + X_{t,I_s}^O }{X_{t,active}^{real}}  - 1\right)^2 
\right],
\end{align}
where $P = (P^{L\to I_m}, P^{I_m\to R}, P^{I_m\to I_s}, P^{I_s \to R}, P_1^{I_s \to D})$ denotes the vector of transition probabilities and $\varphi$ (taken to be 10 \cite{weinberger2020estimation}) is an inflation factor that is used to generate the initial unobserved states from the initial observed real data. That is, for the observed process, we set the initial states by
\begin{align*}
\hat X_{1, I_s}^{O} = X_{1, I_s}^O &= p_S \cdot X_{1, active}^{real}\\
\hat X_{1, I_m}^{O} =X_{1, I_m}^O &= (1-p_S) \cdot X_{1, active}^{real}\\
\hat X_{1, D}^{O} =X_{1, D}^O &=  X_{1, D}^{real}\\
\hat X_{1, R}^{O} =X_{1, R}^O &=  X_{1, R}^{real},
\end{align*}
where $p_S$ is the proportion of severe infections among the active cases. For the population process, the initial states are set to
\begin{align*}
    X_{1,j} &= \varphi \cdot \hat X_{1, j}^O \quad \text{ for } j \in \{ I_m, I_s, R, D\},\\
    X_{1,L} &= 0.5 \cdot X_{1, I_m}.
\end{align*}
Finally, we note that the reason we used ratios instead of differences in the loss is that the numbers in different compartments have different orders of magnitude. For example, the number of infected cases is much larger than the number of deaths. The expectation in the above loss function is computed empirically by taking the average of the loss function over several runs of the processes. 

\textbf{Initial Estimates.} Referring back to Algorithm 1, note that we need $\hat P_{\text{init}}$: initial estimates of the inter-compartmental probabilities to  start the algorithm. These estimates were obtained by matching certain features of the process with the available data. For example, since the time to transition from state $L$ to state $I_m$ is geometric with probability $P^{L\to I_m}$, its expected value will be $1/P_{L\to I_m}$. On the other hand, from studies on the incubation period of the disease\cite{lauer2020incubation}, we known that the average time from latent to mildly infected is $5$ days. Thus,  we obtain the initial estimate $\hat P_{\text{init}}^{L\to I_m} = 0.2$. Similar calculations for other transition probabilities yield:
\begin{align*}
\hat P_{\text{init}}^{I_m \to I_s} &= 0.017, &\hat P_{\text{init}}^{I_m \to R} = 0.024 \\
\hat P_{\text{init}}^{I_s \to R} &= 0.012, &\hat P_{1, \text{init}}^{I_s \to D} = 0.009.
\end{align*}

\subsection{Prediction and Policy Optimization} 

\label{reward_policy}
{\bf Extrapolating beyond day $T$ (prediction):} Given the parameter estimates obtained in the last section, we can now project into the future and optimize the lockdown strategy based on these projections. This is of critical importance as it can serve as a principled guide for real world policy makers.  Consider running the population process for $\tilde T$ days from day 1 (where $\tilde T > T$) for forecasting purposes. We can extrapolate the testing probability for the additional $\tilde T - T$ days in terms of the testing probabilities obtained for the first $T$ days via the following rule: if there is no trend in the testing rates over the first $T$ days one can use the average, whereas if there is a monotone trend, one can use the value at time $T$. 

Note that for the training period $1 \leq t \leq T$, the actions are known and fixed, as they are dictated by real-world measures taken during this period, and therefore the policy optimization only takes place during the test period $T+1 \leq t \leq \tilde{T}$. 

\textbf{Policies and actions.} POMDP models are difficult to optimize in full generality, and as our model has several compartments, and will be run with reasonably large populations, the problem of finding the globally optimal policy via maximizing a value function is essentially computationally intractable. Hence, we confine ourselves to the class of so-called \emph{bang-bang} policies (see \cite{artstein1980discrete}, \cite{angell1976existence}, \cite{hermes1969functional}, \cite{sonneborn1964bang} and references therein)-- impose or lift lockdowns. These are relatively easy to optimize over and allow us to avoid formal belief update methods. In our implementations, these policies can be parametrized by three thresholds, henceforth denoted by $l$, $u_1$ and $u_2$, and a parameter $\theta \in [0,1]$ that produces a linear combination $w = (\theta \cdot X_{t,I_m}^O + (1 - \theta) \cdot X_{t,I_s}^O) / N$ that is compared to the three thresholds. Specifically, at each time $t$ where $t$ is divisible by $14$, 

\begin{itemize}
\item if $w \geq u_2$, the highest level of lockdown is enforced for the next 14 days ($a_{t'}=2$ for $t \leq t' \leq t + 13$),
\item if $u_1 \leq w < u_2$, a lower level of lockdown is enforced ($a_{t'}=1$ for $t \leq t' \leq t + 13$),
\item if $w \leq l$, then all lockdown is released ($a_{t'}=0$ for $t  \leq t' \leq t + 13$),
\item if none of the above holds, i.e. $l < w < u_1$, then the status quo is maintained for the next fourteen days ($a_{t'}=a_{t-1}$ for for $t \leq t' \leq t + 13$).
\end{itemize}
We denote this class of policies by $\Pi_B = \{ (l,u_1,u_2,\theta) : 0 < l < u_1 < u_2 < 1 \text{ and } 0 \leq \theta \leq 1 \}$. Given the policy class, we next explain how to quantify the costs associated with a particular policy.

{\bf Immediate reward function and Policies: }In order to quantify the various costs related to the epidemic and the actions taken, we propose the following immediate (daily) reward function given a fixed policy (that is, given the thresholds $l,u_1,u_2,\theta$). Given that at the end of day $t$ we are at state $X_t$ and propose action $a_t$ [which depends on the threshold parameters as explained above], the immediate reward function [observed at end of day $t+1$ under policy $a_t$] is given by: 
\textcolor{blue}{
\begin{align*}
-r_{X_{t}, X_{t+1}, a_t} (l, u_1, u_2, \theta) &  = C_L (X_{t+1,D} - X_{t,D}) + \left( \frac{a_t}{2}\right) C_{E} + \rho C_{L}\left(X_{t+1,I_s} - \mbox{cap} \right)_+,
\end{align*}}
where $C_E$ is the daily economic cost of lockdown for each particular state obtained via scaling the cost for the US according to GDPs:
\begin{align*}
C_E(\text{state}) = \frac{\text{GDP of state}}{\text{GDP of United States}} \cdot \text{(Daily cost of lockdown for the US in \$million)}. 
\end{align*}

(See Table \ref{costs_table} and Section \ref{section:statewise} for the numeric values of these quantities). $C_L$ is cost of life, the loss incurred due to the death of an individual (see Section \ref{section:statewise}). The factor $\rho < 1$  is introduced to account for the loss due to treatment denial once hospital capacity (taken to be 40\% of hospital beds in the state \cite{aha2020fast})  is exceeded: we take this to be a quarter of the life cost. We then optimize over $l,u_1,u_2, \theta$ [via grid-search] by maximizing the expected sum of the immediate reward functions over the time horizon of interest:
 \begin{align*}
   (l^\star, u_1^\star, u_2^\star, \theta^\star) =  \argmax_{l, u_1, u_2, \theta} \sum_{t=T}^{\tilde{T}-1}\Ex [r_{X_t,X_{t+1},a_t}(l, u_1, u_2, \theta)],
 \end{align*}
 \noindent
with the expectation computed by averaging over several runs of the process. We reiterate that in the definition of $r_{X_t,X_{t+1},a_t}$, the action $a_t$ depends on the policy parameters $l,u_1,u_2, \theta$, and $X_t$ itself depends on the actions taken in the history of the process. Note that since the action $a_t$ assumes a value of $2$ under full lockdown, in the reward function we divide the action by $2$ so that the economic cost under full lockdown is $C_E$ and under partial lockdown $C_E / 2$.

For concreteness, we present the policy optimization procedure in Algorithm \ref{alg:pol_opt}. In practice we replace the class of policies $\Pi_B$ by a discretized version $\Pi_d$.

\begin{algorithm}[H]
\SetAlgoLined
 \textbf{Input: } Fitted parameters $\hat{P}$, The class of policies $\Pi_d$, Number of replicates $J$ \;
 \For { $ (l, u_1, u_2, \theta) \in \Pi_d$}{
 	\For {$j = 1, \dots, J$}{
	Run the population process $X(t)$ for days 1 through $\tilde T$ with estimated parameters $\hat{P}$ and policy $(l, u_1, u_2, \theta)$ to compute
	$R^{(j)}(l, u_1, u_2, \theta) = \sum_{t=T+1}^{\tilde{T}} r_{X_t,X_{t+1},a_t}(l, u_1, u_2, \theta)$\,\;
	}
	Approximate the expected reward by $\bar{R}(l, u_1, u_2, \theta))  = J^{-1} \sum_{j=1}^J R^{(j)}(l, u_1, u_2, \theta) $\,\;
 }
 Find the optimal policy $(l^\star, u_1^\star, u_2^\star, \theta^\star) =  \argmax_{(l, u_1, u_2, \theta) \in \Pi_d} \bar{R}(l, u_1, u_2, \theta))$;
 
\textbf{Output:} Return $(l^\star, u_1^\star, u_2^\star, \theta^\star)$.
 
 \caption{Finding Optimal Policies}
 \label{alg:pol_opt}
\end{algorithm}

\subsection{Sensitivity analysis}
Once the optimal policy for a given location has been determined till time $\tilde T$ one can plot the evolution of the mean curve of the population process from days $T+1$ through $\tilde T$ under the optimal policy for any compartment of interest (e.g. deaths or infected), with compartment probabilities given by the least squares estimates $\hat P$. We next discuss quantifying the sensitivity around such a mean curve. There are two sources of uncertainty: (1) Variability in the process given a fixed set of parameters for its evolution, and (2) Error in estimating the parameters. We combine these two sources of variation to generate a prediction band. Towards that end, we perturb the estimated parameters by adding some noise to the log-odds of these probabilities and transforming back to the original scale. However \emph{we do not change the optimal policy}: we run the population process using this noisy version of the optimal transition probabilities over a certain time horizon ($\tilde{T} - T = 50$ days) multiple times, starting from the end of the training period and obtain the mean curve, which takes care of the first source of variation. We then repeat this experiment multiple times with different sets of noisy parameters which accounts for the second source of variation. Finally, we calculate upper and lower 5th percentiles of the mean curves thus produced (for each time $t$) to generate the sensitivity band. Our algorithm can be summarized as follows: 
\\

\begin{algorithm}[H]
\SetAlgoLined
 \textbf{Input: }The estimated parameters $\hat P$, The (simulated) state of epidemic $X_T$ at the end of training period, generated under $\hat P$ \;
 \textbf{Outer loop iterations: }Set number of outer loop iterations = $O$ \; 
 \textbf{Inner loop iterations: }Set number of inner loop iterations = $I$\;
 \textbf{Initialization: }Set the variables $x = y = 1$\;
  Set $\ell_j = \log(\hat{P}_j / (1 - \hat{P}_j))$ for $j=1,\dots,5$.\;
 \While{x  $ \leq O$}{
    Generate  $\tilde{\ell}_j \sim \mathcal{N}(\mu = \ell_j, \sigma = \ell_j / 3)$ for $j = 1, \dots, 5$\;
    Set $\hat P_j^{now} =(1 + \exp( -\tilde{\ell}_j))^{-1}$ \;
  \While{y $ \leq I$}{
   Generate the population process using $\hat P^{now}$ starting at $X_T$ and keeping the threshold parameters (determining the actions) at their optimized values. 
    }{
   Store the mean curve of the output curves from the inner loop. 
  }
 }
 \vspace{0.2cm}
 \textbf{Quantile selection: } For each day, select the $5\%$ and $95\%$ percentile value of the $O$ many mean curves\;
 \vspace{0.2cm}
 \textbf{Upper band: }The curve traced out by the $O$ many $95\%$ quantiles \; 
  \vspace{0.2cm}
 \textbf{Lower curve: } The curve traced out by the $O$ many $5\%$ quantiles. 
 \caption{Getting the uncertainty band}
\end{algorithm}

The perturbations are in the log-odds scale to avoid boundary issues involving the inter-compartment probabilities, as many of them are small numbers. The choice of a standard deviation for the noise of the same order as the mean but smaller than it allows exploration of parameters that are on the same scale as $\hat{P}$, but not too far from it.

\subsection{Statewise Numerical Analysis }\label{section:statewise}
%In order to make the problem more tractable, we focus on a class of simple policies known as bang-bang policies in the control theory literature. For each county, these policies enforce (or relax, respectively) a lockdown depending on whether the convex combination $T_t := \frac{\theta X_t^O(\text{mild} + (1 - \theta) X_t^O(\text{severe})}{N}$  crosses an upper threshold $u$ (or a lower threshold $l$, respectively). In other words, a lockdown is enforced whenever $T_t$ moves beyond $u$, and it is kept in place until the next time $T_t$ crosses the lower threshold $l$.
%\begin{figure}[H]
%	\includegraphics[scale = 2, width=\linewidth]{mdp_c1.png}
%	\caption{The effect of lockdown on the (daily) number of infected in county 1. Shaded areas show the days lockdown was enforced. Each lockdown period lasts for 42 days. Flattening the curve results in a $17\%$ decrease in the number of deaths.}
%	\label{mdp_plots}
%\end{figure}
In this section we provide detailed analysis of six states: Michigan (MI), New Jersey (NJ), Florida (FL), Arizona (AZ), Texas (TX) and California (CA). %For the northern states MI and NJ, where the epidemic peaked early, real data (consisting of daily active cases, recovered and deaths) from May 1 to May 31 were used to estimate the model parameters, whilst for the others data from June 1 to June 30 %for southern states is used to estimate model parameters (transition probabilities) using the least squares criterion \ref{eq:ls_criterion}.
%were deployed. 
For each state, optimal values of the inter-compartmental transition probabilities were obtained by analyzing data from its training period. Given these estimates from the respective training periods, policy optimization was accomplished as described in Section \ref{reward_policy}. Recall from subsection \ref{reward_policy}, that we need both the cost of lockdown $C_E$ and the cost of life $C_L$ to obtain the optimal policy. These were calculated in the following way: 
\begin{enumerate}
\item For $C_E$, we started with a suggested \$7 trillion per year for the United States \cite{casey2020some}, that is, roughly \$20 billion per day for all the states combined. Then we scaled this number by the fraction of the GDP of each state to the US GDP to obtain the economic cost (under lockdown) to the GDP for the state of interest. These numbers are displayed in the second column of Table \ref{costs_table}. 
\item For the cost of life $C_L$ we use \cite{conover2020how}, where \$4.7 million is given as value of statistical life year (VSLY) and a lower value of \$1.5 million is obtained as the equal-value life year gained (see \cite[Figure 2]{conover2020how} for a comparison of these estimates). We ran our simulation under both these costs. 
\end{enumerate}
\begin{table}[]
\center
\begin{tabular}{|llll|}
\hline
 &State  & Econ. cost & Medical capacity \cite{aha2020fast}  \\
 & & (Per day in millions) & (No. of  beds for Covid patients) \\ \hline 
 & AZ & 341.8 & 6537 \\
 & CA & 2928 & 34242 \\
 & FL & 978.6 & 24208 \\
 & MI & 484.7 & 11320  \\
 & NJ & 577.1 & 10398  \\
 & TX & 1761 & 30784  \\ \hline
\end{tabular}
\caption{Economic cost of lockdown and medical capacity. The number of available hospital beds for Covid patients at each state is taken to be $40\%$ of all available beds \cite{aha2020fast}.  }
\label{costs_table}
\end{table}
%Of the six states analyzed in this paper, three (MI, TX, CA) are presented in the main body of the paper and rest are in supplement. 
The six states were divided into two broad categories: \emph{northern states} (MI, NJ) and \emph{southern states} (TX, AZ, CA and FL). For each of the states, we trained using real data from 40 days, then predicted the process for the next 50 days (thus, a total 90 days starting from the start of the training period). For northern states our training data starts from May 1 and for the southern states from June 1, the reason for this discrepancy being the recent surge of covid cases in the southern states, which renders the data from May less informative for prediction purposes at this time. We present MI, TX and FL in the main body of the paper and the remaining three states in the supplement. 

\begin{figure}[t!] % "[t!]" placement specifier just for this example
\begin{subfigure}{0.48\textwidth}
\includegraphics[width=\linewidth]{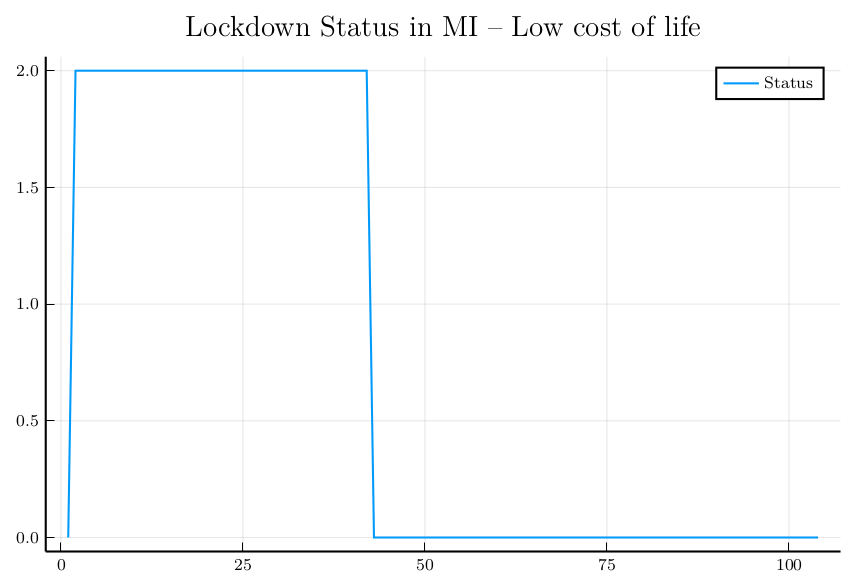}
\caption{} \label{fig:a}
\end{subfigure}\hspace*{\fill}
\begin{subfigure}{0.48\textwidth}
\includegraphics[width=\linewidth]{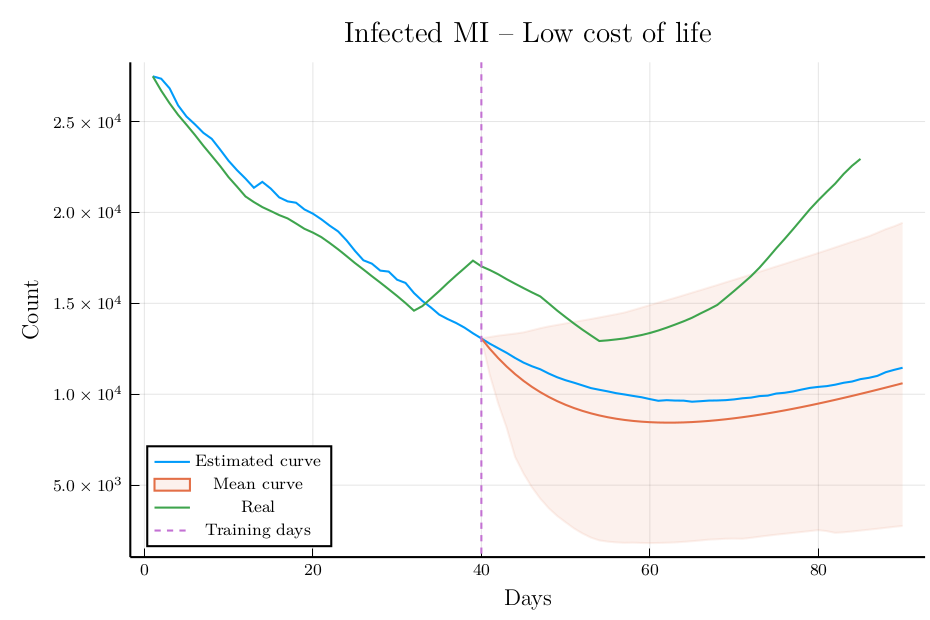}
\caption{} \label{fig:b}
\end{subfigure}

\medskip
\begin{subfigure}{0.48\textwidth}
\includegraphics[width=\linewidth]{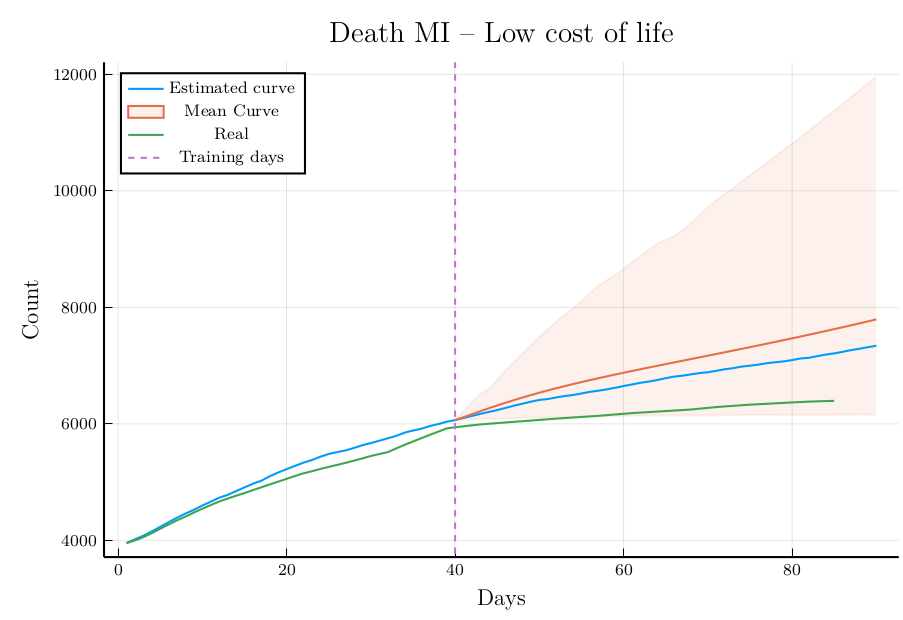}
\caption{} \label{fig:c}
\end{subfigure}\hspace*{\fill}
\begin{subfigure}{0.48\textwidth}
\includegraphics[width=\linewidth]{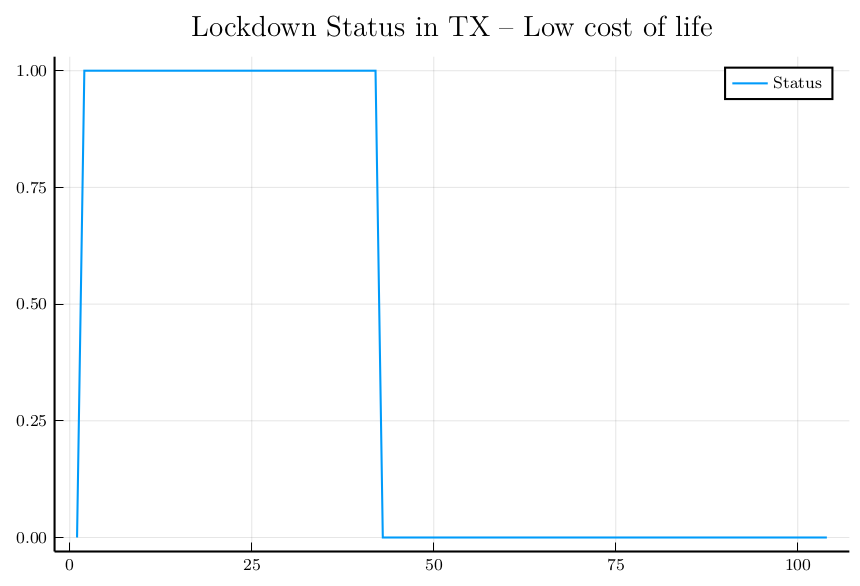}
\caption{} \label{fig:d}
\end{subfigure}

\medskip
\begin{subfigure}{0.48\textwidth}
\includegraphics[width=\linewidth]{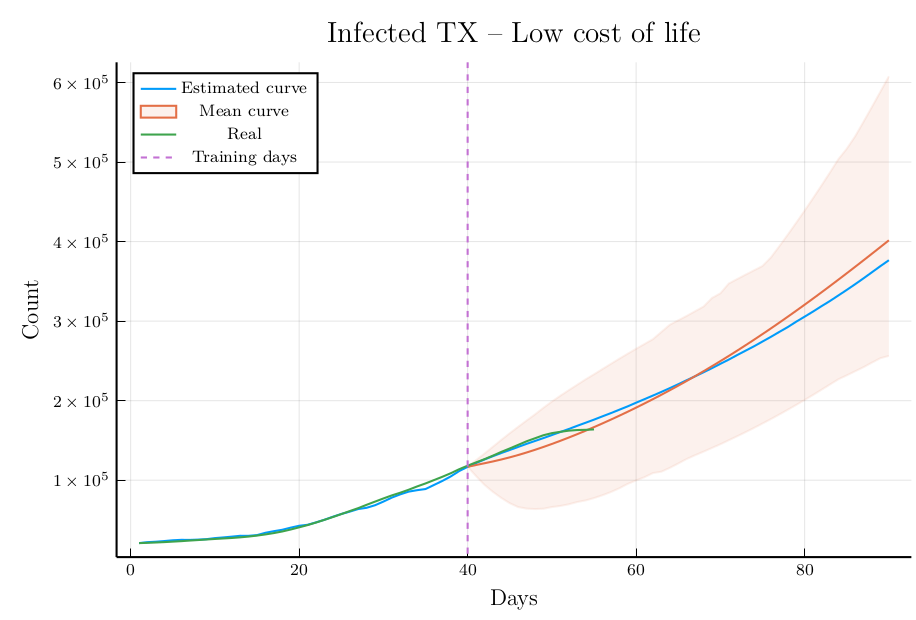}
\caption{} \label{fig:e}
\end{subfigure}\hspace*{\fill}
\begin{subfigure}{0.48\textwidth}
\includegraphics[width=\linewidth]{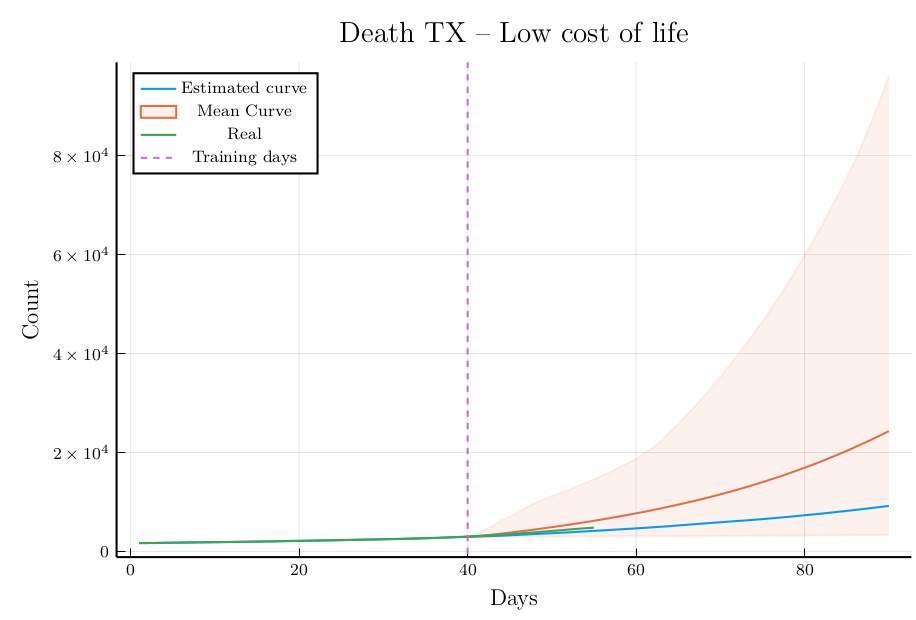}
\caption{} \label{fig:f}
\end{subfigure}

\caption{Analysis of MI and TX} \label{fig:1}
\end{figure}

\begin{figure}
\begin{subfigure}{0.48\textwidth}
\includegraphics[width=\linewidth]{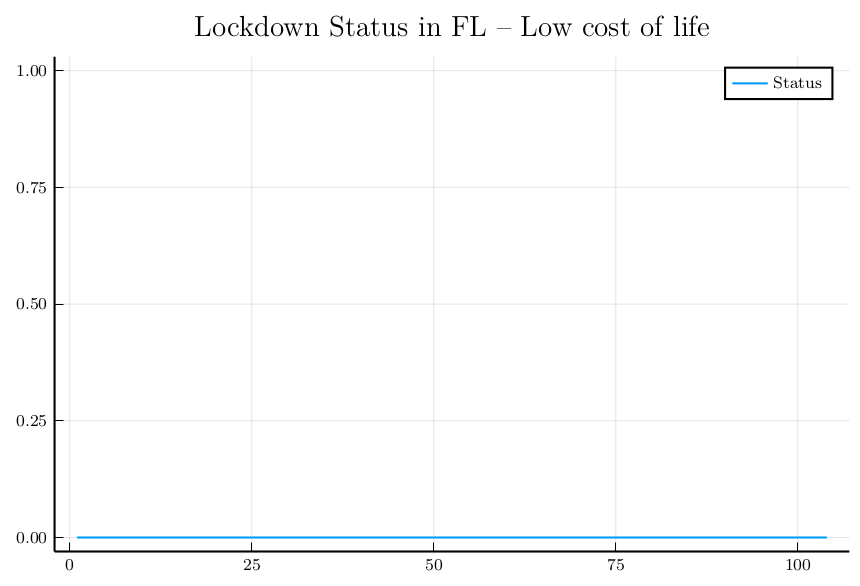}
\caption{} \label{fig:d}
\end{subfigure}
\medskip
\begin{subfigure}{0.48\textwidth}
\includegraphics[width=\linewidth]{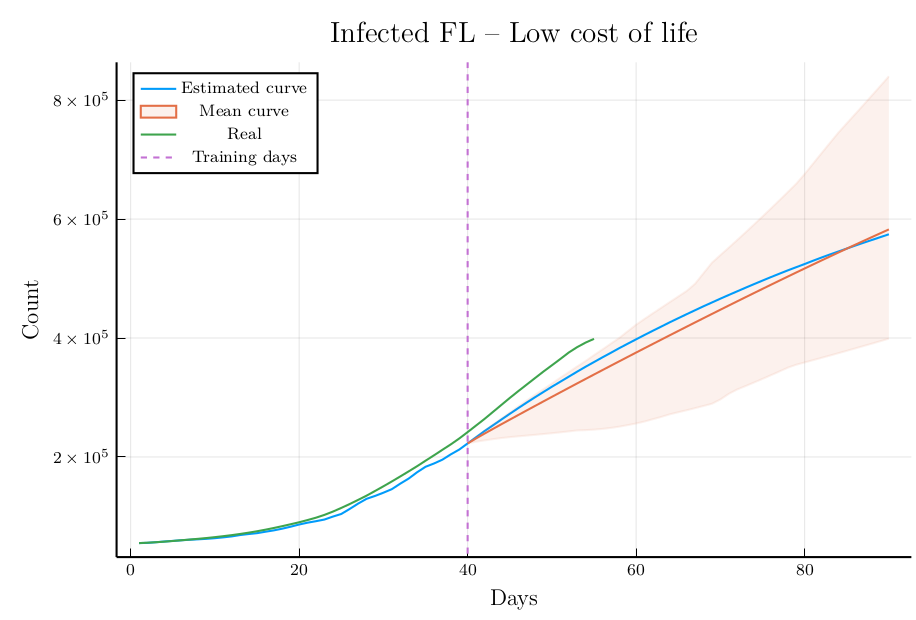}
\caption{} \label{fig:e}
\end{subfigure}\hspace*{\fill}

\begin{subfigure}{0.48\textwidth}
\includegraphics[width=\linewidth]{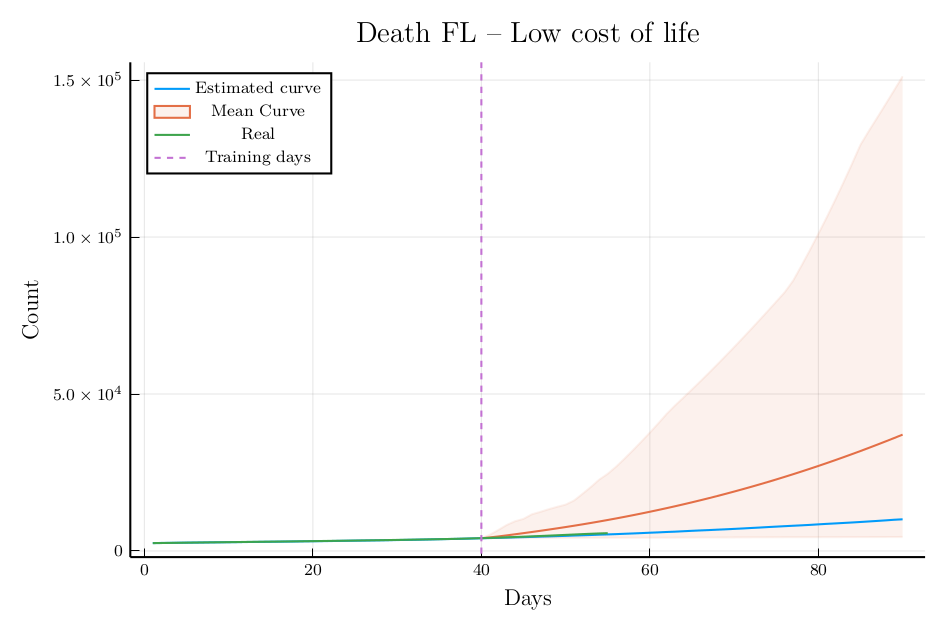}
\caption{} \label{fig:f}
\end{subfigure}

\caption{Analysis of FL}
\end{figure}

\begin{figure}
\begin{subfigure}{0.48\textwidth}
\includegraphics[width=\linewidth]{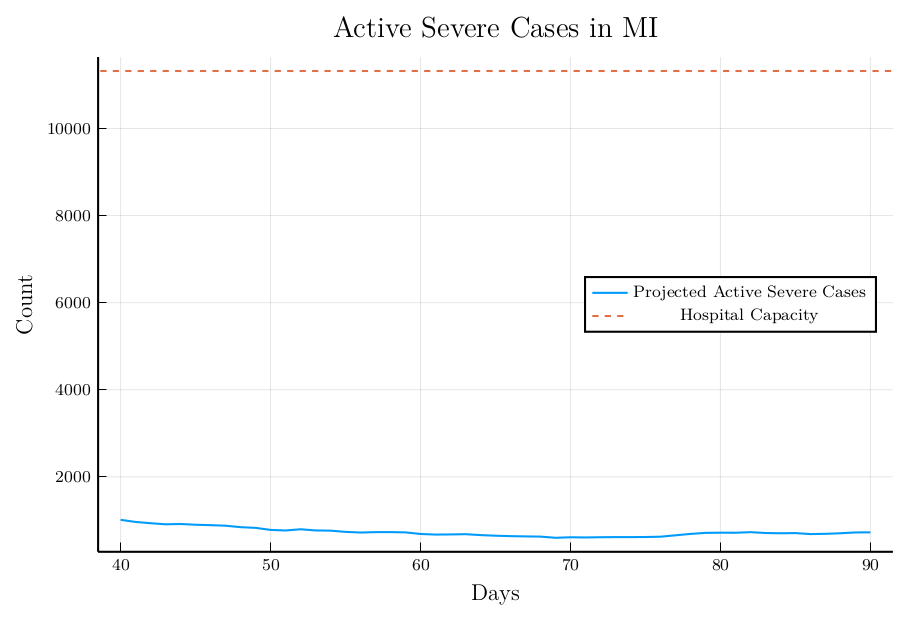}
\caption{} \label{fig:d}
\end{subfigure}
\medskip
\begin{subfigure}{0.48\textwidth}
\includegraphics[width=\linewidth]{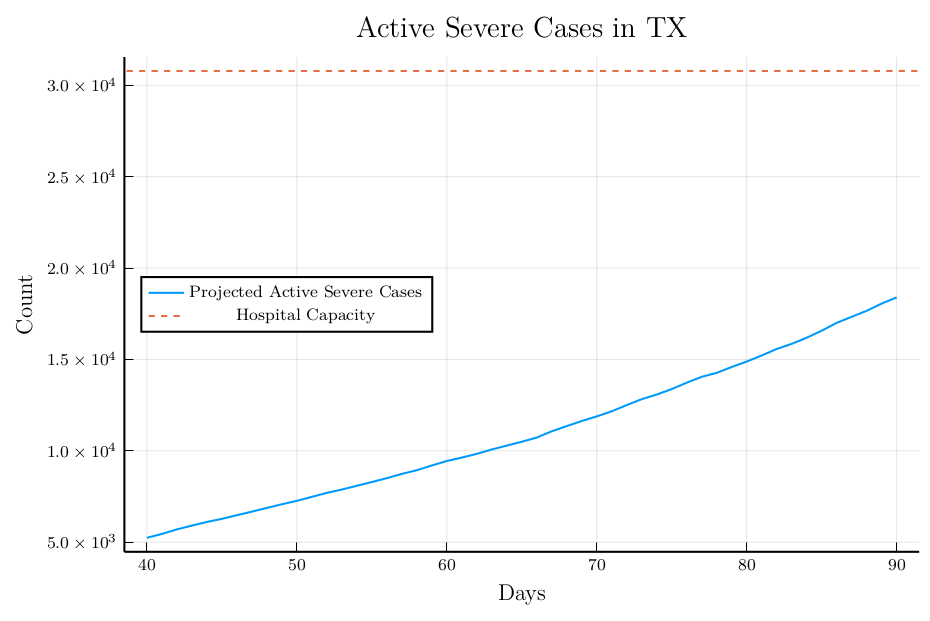}
\caption{} \label{fig:e}
\end{subfigure}\hspace*{\fill}

\begin{subfigure}{0.48\textwidth}
\includegraphics[width=\linewidth]{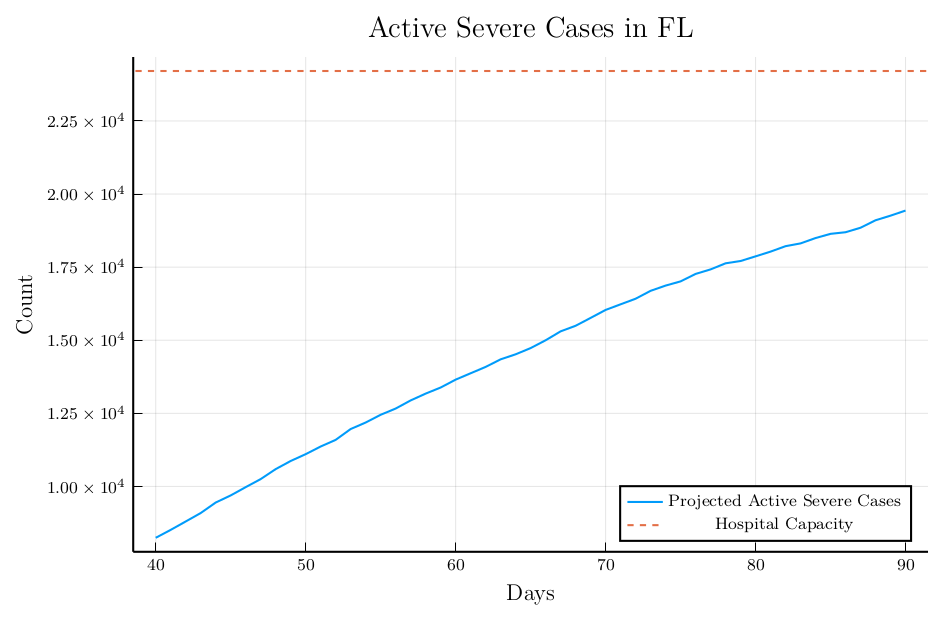}
\caption{} \label{fig:f}
\end{subfigure}

\caption{Severe Cases vs Hospital Capacity}
\end{figure}

For each of three states (MI, FL, TX) we present three pictures: One for the time-varying policy, i.e. lockdown status over 90 days, the second showing the number of people in the infected compartment (sum of mildly ill and severely ill people) and the third exhibiting the number of deaths. Recall that lockdown status is coded as 2 for full lockdown, 1 for partial lockdown and 0 for no lockdown. For MI (and NJ, see supplement), we work under a full lockdown status over the training period [since both these states were under strict controls during this time] and for TX (as well as CA and AZ, see supplement) we work under partial lockdown over the training period [these states had a number of restrictions in place but many activities and businesses were still functional] and optimize the policy for the next 50 days. FL was trained under no lockdown in keeping with its status during the training phase. In the pictures corresponding to the infected and death compartments, we display 3 curves: the blue curve corresponds to the evolution of the process using $\hat P$ (the least squares estimate from Algorithm 1) labeled as \emph{Estimated Curve}, the green curve labeled as \emph{Real} shows the actual observed numbers for that state and the \emph{Mean Curve} in orange is the mean of the trajectories of the process using the perturbed parameters. More precisely, as described in Algorithm 4, we obtain a mean process for each set of perturbed parameters and the \emph{Mean Curve} is the just the mean of those mean curves. The shaded region represents the $90\%$ band based on the mean curves generated by the perturbed parameters as described previously. The dashed vertical line separates the training period from the test period.  

For $p_S$ we use (for all states but NJ) an estimated value of $0.07$ which is approximately the proportion of daily active cases that are hospitalized. This quantity generally ranges between 5 and 8 during the training period for all the states we consider other than NJ \cite{jhu2020data,ken2020tracking,fl2020dashboard}. For NJ we use an estimated value of $0.15$ \cite{nj2020tracking}.

We performed our analysis twice for each state, once for high life cost and once for low life cost, and in each case we obtained the same optimal policy: lift the lock down (i.e. $a(t) = 0$ after training period). This may seem surprising at first sight given that the numbers are rising fast especially in the southern states but a closer inspection of the available data provides some justification for the prescribed optimal policy. 

Consider the the projected numbers of severely infected and the figures on the number of hospital beds allocated for covid patients in Table \ref{severe_table}. It is projected that MI will have in the ballpark of 700 severe cases by the end of July.\footnote{At the time of writing at the end of July, MI has 438 hospitalized cases, which suggests our prediction is over-estimating even if we account for some unobserved severe cases. On the other hand, at the end of July, NJ reported about 700 hospitalized cases, which is higher than our projection of 501. (Similar data for other states, corresponding to the end of August, is not available at the time of writing.)} On the other hand MI has around 11000 hospital beds for Covid patients, which is clearly more than sufficient. While the projected number of deaths in Michigan shown by the blue curve in Figure (c) over the test period [till July 25th] is over-estimating the real death curve [orange] which is quite flat, the divergence is still modest, and the $a(t)=0$ policy over the training period is quite compatible with the slow growth rate of both projected and real death curves. As far as Texas and Florida are concerned, unlike Michigan (and New Jersey, shown in the supplement), the number of active severe cases is projected to increase through the end of August, in keeping with the growing phase of the epidemic at present, but the growth of severe cases is not particularly fast: for FL, AZ the projected curves are concave whereas for TX and CA they have a roughly linear trend; at any rate there is no `shooting-up' that might cause grave alarm. As for MI, the projected number of severe cases stays well below hospital capacity, i.e. the `curve stays flat'. In summary, for each state under consideration, given the available number of (COVID) hospitable beds, the number of people who currently are (and predicted to be in the future) severely ill, and the current and predicted numbers of deaths, the policy finds the economic cost incurred due to lockdown to be too high from the optimization point of view. It is also evident from our picture, especially from the test region, that our model has overall decent predictions and the projected numbers lie inside the band in most cases. % An inspection of the number of deaths in Michigan from June onwards [recall that the test period for Michigan is June 10 through end of July] shows an in-control decreasing tend, showing that the optimal policy (based on the earlier training data) has made a reasonable call. 
%Similarly, the number of severe cases is projected at approximately 15000 at the end of August, which is well bellow the hospital capacity taken to be more than 24000. 
Two further points need to be noted. First, the presented bands are not based on a strict inferential principle and must be interpreted as a rough exploration of variability (sensitivity to perturbations of the least squares estimated) more than anything else. As far as the numbers of infected go, the bands generally tend to be overly conservative on the lower end, whereas in terms of number of deaths one notices just the opposite trend: the upper bounds are ultra conservative now. More insights into the performance of these bands is needed. Second, the policy $a(t)=0$ over the test period should be interpreted with caution. Recall that for us $a(t)=0$ still corresponds to an $R_0$ value that is appreciably smaller than the values around 2.5 or more that were estimated in the pre-epidemic phase: in other words, no lockdown is \emph{not equated to} a return to normal pre-pandemic life, but public health informed behavior with adequate distancing whenever possible and use of masks while generally going about one's business. Most activities and occupations can continue under such circumstances. See also the caveats we discuss in Section \ref{sec:extension} in connection with the interpretation of prescribed policies.

\begin{table}[]
\center
\begin{tabular}{|llll|}
\hline
 State  & Active Severe Cases& Hospital Capacity & Number of Deaths  \\\hline
  AZ & 4599 & 6537  & 5389 \\
  CA & 15617 & 34242 & 12386 \\
  FL & 19435 & 24208 &  10086 \\
  MI & 725 & 11320 & 7342  \\
  NJ & 501 & 10398 & 14029  \\
  TX & 18396 & 30784 & 9216  \\\hline
\end{tabular}
\caption{The number of active severe cases and the cumulative number of deaths predicted on day $\tilde{T} = 90$.}
\label{severe_table}
\end{table}

\newpage
\section{Non-Markovian Model}
\vspace{-1.5cm}
\label{sec:Non_Markovian}
\begin{center}
\begin{figure}
\includegraphics[width=1\linewidth]{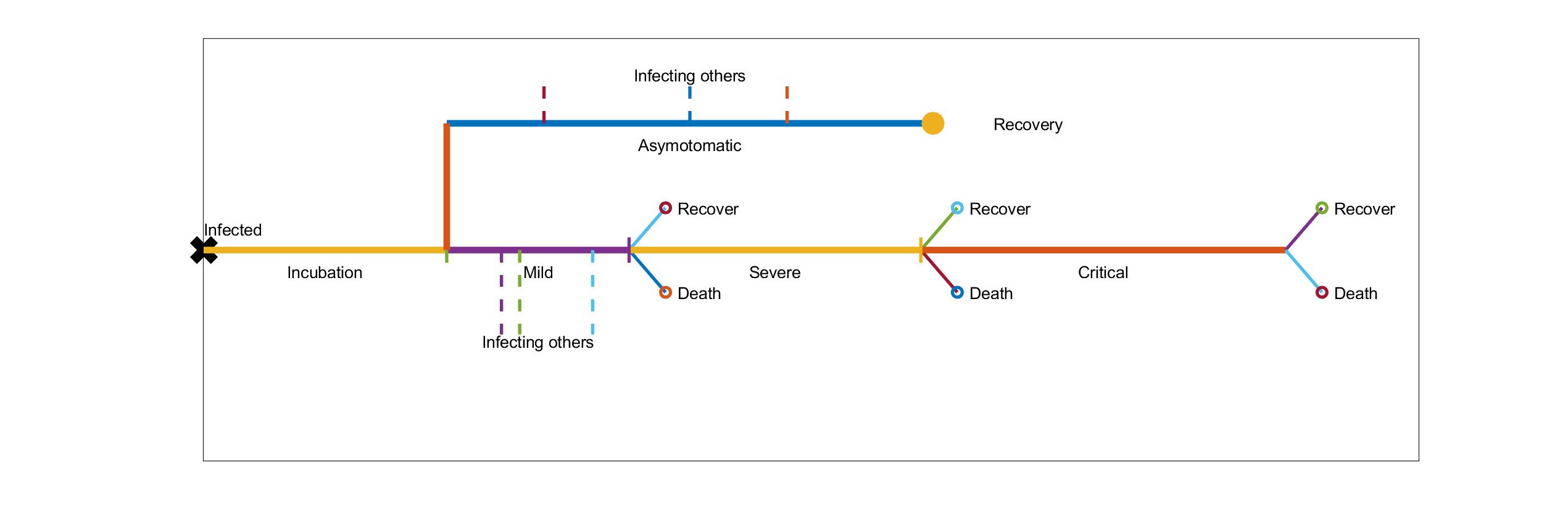}
\caption{A disease history of a patient}
\label{nmph}
\end{figure}
\end{center}
The vast majority of existing literature on pandemic modeling focuses on Markovian processes, not least owing to a broader understanding of decision making strategies in such frameworks and optimization techniques thereof, as well as the existence of ready software. While non-Markovian processes have been investigated in epidemic propagation contexts, and mostly so in network based analysis (see \cite{pang2020functional}, \cite{sherborne2018mean}, \cite{yang1972empirical}, \cite{vizi2017pairwise} and references therein), the literature in this arena is significantly sparser, and very few works in the recent slew of papers on the pandemic have considered this angle. However, we believe that exploration of non-Markovian models is quite important since the standard Markovian compartment models used in practice do not quite describe epidemic transmission accurately. While they provide good working approximations, they are not good at capturing natural delays, and accounting for the fact that the probability of an individual transitioning from one compartment (say $I_m$)  to another (say $R$) typically depends on their history in the $I_m$ compartment and is actually time-dependent. In the case of COVID, for example, the delay between being acquiring the infection ($L$) to being able to transmit the virus ($I_m$) has a mode at 5-7 days \cite{lauer2020incubation}, while the delay between being the $I_m$ state and death has a mode of around 3 weeks [see \href{https://www.hopkinsguides.com/hopkins/view/Johns_Hopkins_ABX_Guide/540747/all/Coronavirus_COVID_19__SARS_CoV_2}{here}]. The parsimonious Markovian model is restricted to geometric delays with a mode at 0: under a geometric assumption, the memoryless property guarantees the Markovian property with a time-homogeneous transition probability (for i.i.d. individual trajectories). It is, of course, possible to extend the model by including compartments for each day between infection and disease: e.g., from the compartment of 3 days after acquiring the infection, a subject moves either to the compartment of 1st day of disease or to the compartment of 4th day of infection. But this becomes exceedingly cumbersome.  
Another issue to keep in mind is that in a decision based framework, the effect of imposing a lockdown typically influences infection rates \emph{between households} but not \emph{within}, and therefore a reduction in transmission propensity uniformly over all individuals may not conform to a good modeling strategy especially in locations with typically large households. Also, the effect of imposing a lockdown is not immediate, there are natural delays in its coming into effect, so to speak. It is considerably easier to accommodate these diverse issues in a non-Markovian framework. 
%for example, the delay between the time of maximal new infections per day to the day of maximal mortality. A more natural model for the epidemic would be non-Markovian like the model we describe next.
Towards that end, we next present some ideas for a more flexible non-Markovian model which, instead of modeling the compartments, models each person separately. The development at this point is preliminary, however as these ideas bear promise and point to new paradigms for studying future epidemics, we deem it worth a discussion. 

Consider a population of $N$ members (for our current simulation $N= 5 \times 10^6$, from which some results are displayed below) who are located in $K$ ($=9$ in the current simulation) clusters, each further divided into communities. For better understanding, one may think clusters as states and communities as counties [or on a shorter scale, clusters as counties and communities as various municipalities within the county].  The population is divided into two sub-groups based on their risk of severe illness upon infection: each member $i$ has an independent risk factor $\rho_i$, primarily governed by his/her medical profile and age, with $\rho_i>r$ considered high-risk, and low risk otherwise. An infected member $i$ can potentially infect $I_i \sim \text{Poi}(\lambda)$ members, with $\lambda = R_0$. The natural history of a typical subject after being infected is described in Figure \ref{nmph}. After the incubation period ($L$), a subject moves either to be ``asymptomatic'' or ``mildly'' ill (defined as a carrier that is not hospitalized). During these two periods the subject may infect others. From the asymptomatic state, the subject recovers after a few days, while from the mild illness state, the subject may transit to recovery, die without hospitalization or become ``severely ill'' (i.e.  is hospitalized and recorded as being severely ill). By the end of this interval, she may die, recover or become ``critically ill'' (i.e., transferred to the ICU). Finally, after the ICU period, she may recover or die.  

The probability of a susceptible picking up the infection depends on the social-distancing policy enforced at the time of the potential infection, since that influences $R_0$. Once infected, the transit probabilities  between the different stages depend on the risk factors, e.g. a person in the high risk group will be more likely to be in the severely ill compartment, whereas, a person in the low risk group will either recover from the asymptomatic stage or the mild infection stage with high probability. The period of time a person is capable of infecting others is taken to be i.i.d. (the common distribution was taken as a 
Beta (3,6) distribution scaled to the maximum duration of the mild/asymptomatic intervals for our simulations) across different individuals. Denote by Inf$(j)$, the set of individuals infected by an person $j$. We pick every member of Inf$(j)$ from a mixture distribution such that with high probability they are in close proximity to $j$ [i.e. in the same community] and with small probability at a substantial distance [other communities and clusters]. Define by $H_{kt}$ (respectively $L_{kt}$) the number of active cases at time $t$ in cluster $k$ who are at high risk (respectively low risk). An individual $j$, at the end of their latency plus disease period, dies with probability $q(\rho_j,\sum_k H_{kt})$ on day $t$ for some function $q$ (or recovers otherwise). The function $q$ is chosen so that it enables modeling the chance of dying in terms of the person's individual risk as well as the available medical facilities (which would be typically overburdened for high values of the sum in the second argument). 

The process starts with $N_0$ newly infected members. The current implementation enables importing a small number of cases from outside (outside of the clusters considered) daily. The following social-distancing policy is simulated: If for some function $h$, $h(L_{kt},H_{kt})<c_1$, social-distance restriction is relaxed. If $h(L_{kt},H_{kt})>c_2$ social distancing is enforced on high-risk members of cluster $k$, and if also
$L_{kt}/H_{kt}>c_3$, it is applied to all members of the cluster. We take $h$ in our simulations to be a convex combination of the proportions of low risk and and high risk people. The constants $c_1,c_2,c_3$ are optimized to minimize a loss function which depends on the number of days each of the social distancing policies is applied (economic cost), the total number of deaths and the number of hospitalization days of severely ill cases, similar to the loss function we used in the Markovian framework. 
%We are currently running simulations based on different choices of $q$ and $h$. We note that this model allows easy incorporation of longer lag effects by eschewing a Markovian mechanism, which mirrors realistic propagation modes more accurately. 
Figures \ref{fig:A} - \ref{fig:D} represent one such simulation result.
\begin{figure}[H]
\begin{subfigure}{\textwidth}
\centering
\includegraphics[width=1\linewidth]{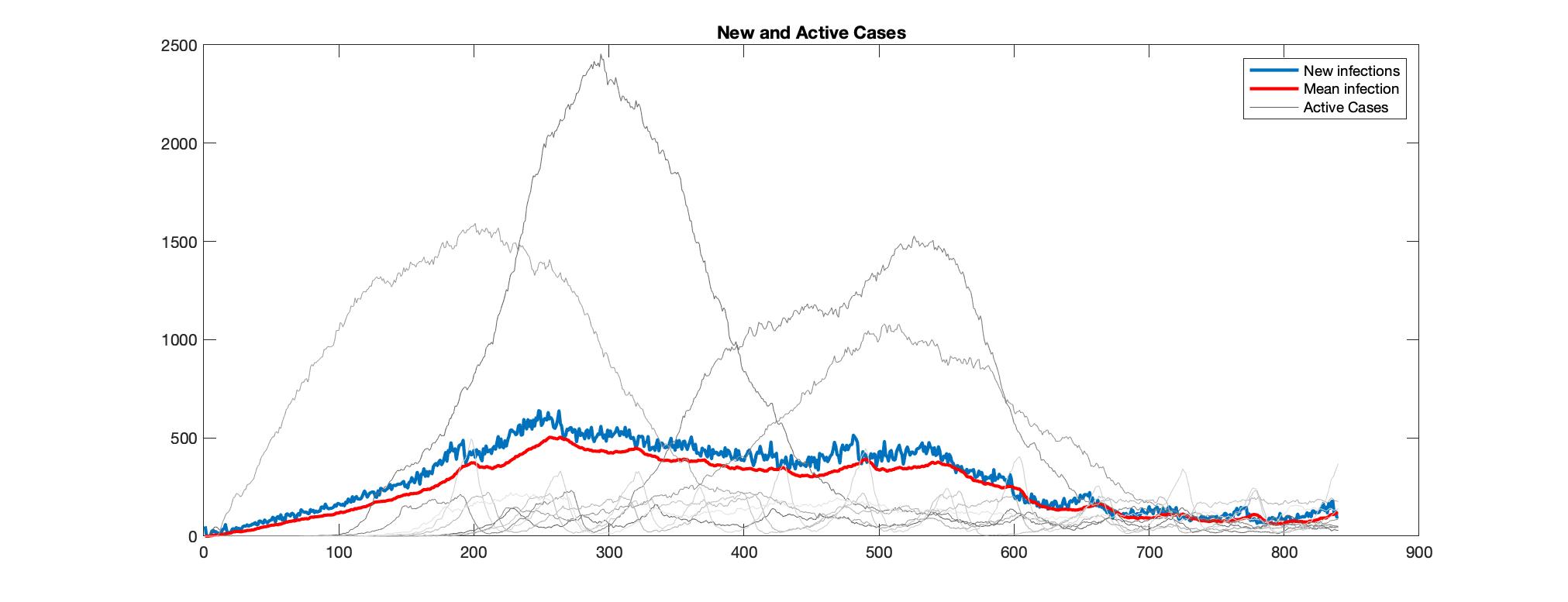}
\caption{Daily cases in different clusters and the average over clusters}
\label{fig:A}
\end{subfigure}
\end{figure}

\begin{figure}[H]
\ContinuedFloat
\begin{subfigure}{\textwidth}
\centering
\includegraphics[width=1\linewidth]{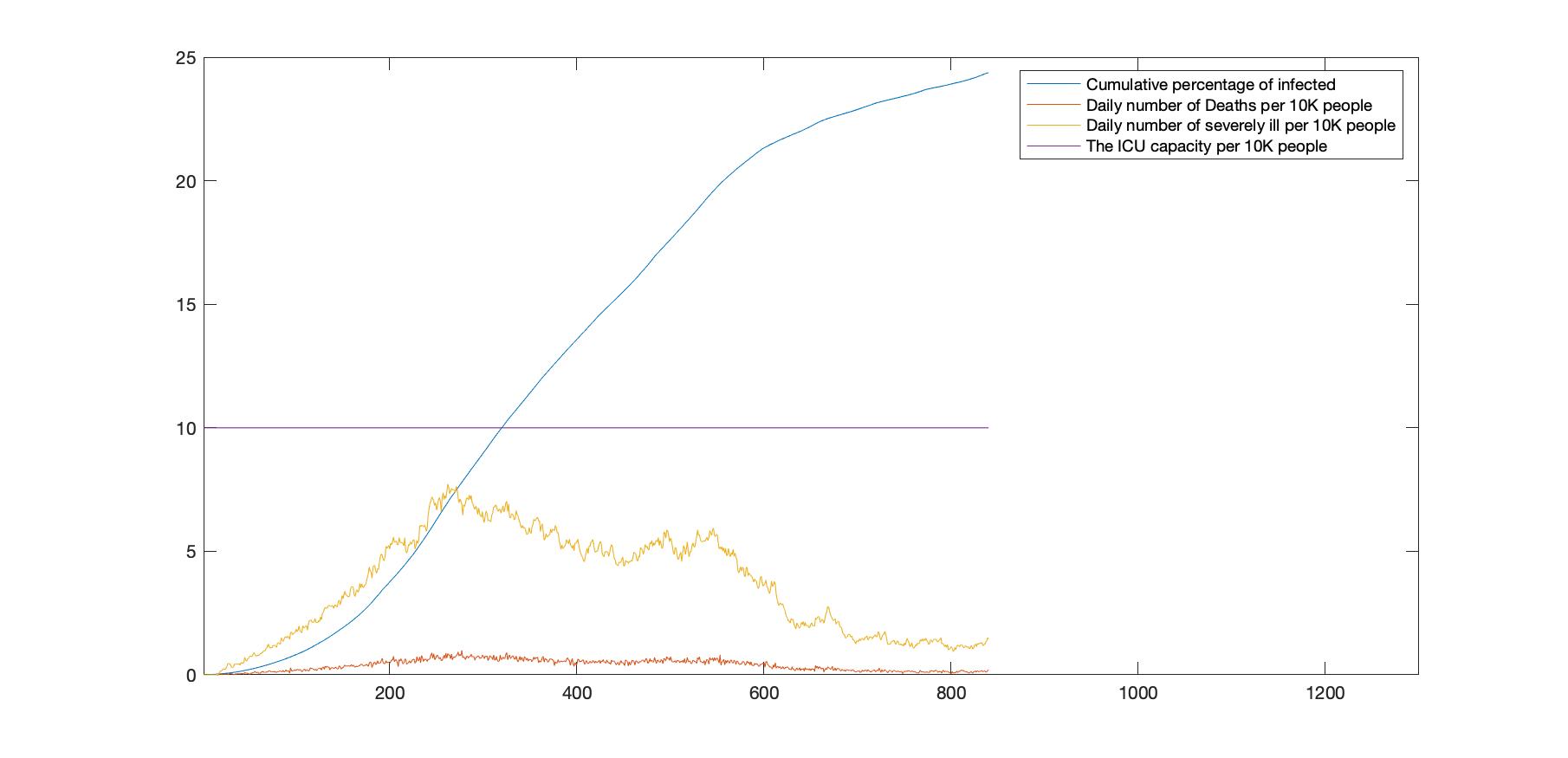}
\caption{Various pandemic numbers over time}
\label{fig:B}
\end{subfigure}
\end{figure}

\begin{figure}[H]
\ContinuedFloat
\begin{subfigure}{\textwidth}
\centering
\includegraphics[width=1\linewidth]{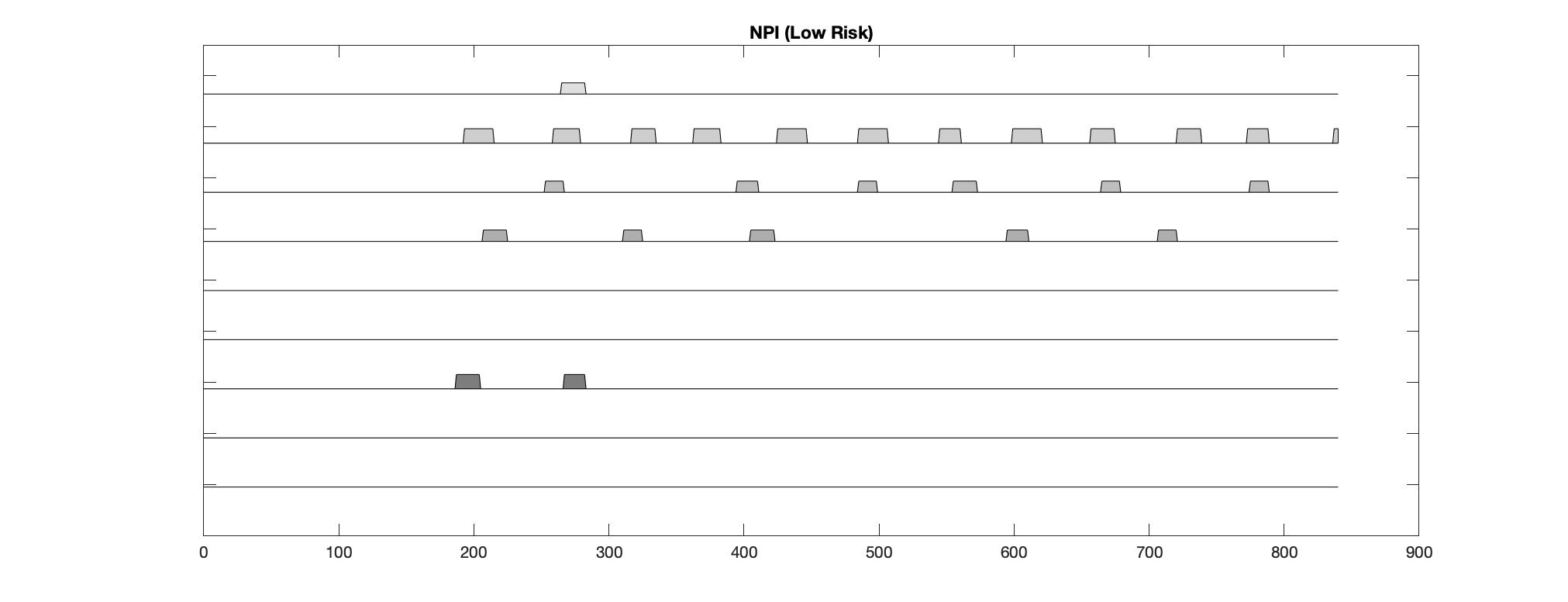}
\caption{Lockdown enforced on low risk individuals by cluster}
\label{fig:C}
\end{subfigure}
\end{figure}

\begin{figure}[H]
\ContinuedFloat
\begin{subfigure}{\textwidth}
\centering
\includegraphics[width=1\linewidth]{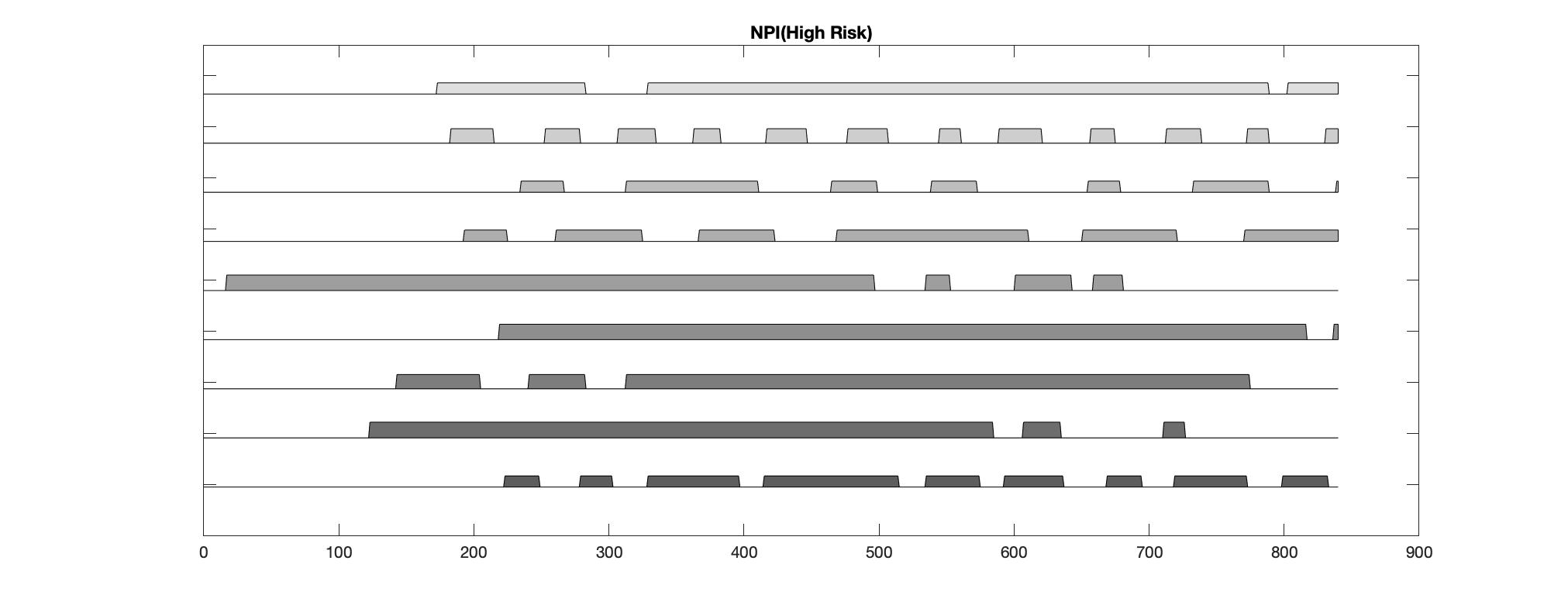}
\caption{Lockdown enforced on high risk individuals by cluster}
\label{fig:D}
\end{subfigure}
\caption{Plots of the Non-Markovian model: For plot (a), $X$-axis denotes the time horizon and $Y$-axis denotes the no. of cases. In (b) we plot daily deaths (red), severely ill (yellow) and ICU capacity (magenta) per 10K people. The blue line in (b) denotes the cumulative proportion of infected people over time. In plots (c) and (d) the $Y$-axis represents clusters and $X$-axis represents the no. of days. The shaded area denotes time duration when lockdown was enforced. The optimal lockdown policies conform to the intuition that frequent and more extensive lockdowns on high risk people are essential. The different shades denoting the lockdowns simply conform to different clusters and carry no other meaning.}
\label{fig:fig_jr}
\end{figure}

\section{Extensions and Future Work}
\label{sec:extension}
\vspace{-.1in}
In this paper we have presented a study of Covid-19 evolution via an SEIRD model within the broad framework of optimal decision making (in terms of lockdown policies) and provided, what appear to be reasonable policies for the states under consideration. We have also sketched some ideas for future development of non-Markovian epidemic propagation models that show considerable promise for the future. In our concluding section, we point out possible extensions and future work in this direction, along with certain caveats of the current approach. 

\subsection{Possible extensions }  Our Markovian model was developed in the framework of optimal bang-bang control policies. One important extension is to allow more complex policies into our set-up, but this requires more involved computation (see \cite{aastrom1965optimal}, \cite{roy2005finding}, \cite{hansen2013solving}, \cite{szer2012maa} and \cite{pineau2003point} and references therein). Typically, lockdowns are often imposed in a phased fashion with essential goods and services phased out last and also made operational first. Such nuances would require introducing more fine-tuned lockdown statuses than the ones we have used in this paper. Furthermore, it may also be of importance to develop more fine-tuned loss functions using input from economists to calibrate the losses due to different phases of lockdown, i.e. differentiate more carefully between the lockdown costs of essential services and goods and their non-essential versions. As of now, we operate under the assumption that the economic impact of a partial lockdown is half of that of full lockdown, but this factor itself needs more investigation.  An important feature of our non-Markovian model is the incorporation of a risk score for each individual in terms of demographic covariates, which allows the enforcement of movement or social distancing restrictions on people depending on their age, health status and medical conditions. This is, in fact, one of the key issues with COVID, as there is enough evidence that the virus can be extremely harsh on older and/or unwell people. We note that a similar distinction between high and low risk individuals can also be incorporated in the Markovian formulation by subdividing each compartment into high risk and low risk compartments, and indeed, our initial strategy for data analysis was to take this into account. Unfortunately, the available data sources do not provide enough information to subdivide these compartments reliably, and we chose to adopt the more parsimonious approach. However, future extensions of this type, should more data become available, would be useful. Finally, the Markovian model we used is purely temporal in the sense that we do not consider spatial effects on transmission of infection, which can sometimes be quite informative. This is not particularly difficult to take into account by sub-dividing a state into sub-regions and using a connectivity matrix to quantify the chance of a random person in a particular sub-region (say, a county) traveling to another sub-region (or staying put), and using this matrix to calculate the chance of a susceptible person picking up the infection by an extension of the current formula at the beginning of Section 2.2. 

%Furthermore, one simplifying assumption in our Markovian model is that all seriously ill cases are known, as they have presented themselves at the hospital. However, as pointed out earlier, at least in the US context, that a non-ignorable fraction of serious cases are being missed, so one important extension will be to divide the S compartment into two sub-compartments: observed and unobserved, and model their trajectories differently.
% We will extend our POMDP model to take into account risk factors and differential policies based on risk. This is not particularly difficult, as all that it requires is sub-dividing the population in each location into varying risk sub-compartments and modeling their transitions separately. We will also explore policy optimization in the formal belief update paradigm.
As far as the non-Markovian approach is concerned,  we note that the model advocated above is flexible enough to incorporate different types of testing (both PCR and serum), more flexible social distancing and social behavior (e.g. networks of members), and different levels of identifiability (e.g., the effect of asymptomatic patients), which could be considered carefully in future work.  Furthemore, modeling individual level variability as opposed to a compartment as a whole, allows a considerable amount of heterogeneity into the framework. The non-Markovian model however suffers from the cons of being computationally more demanding.  In order to make a case for broader adoption of this approach, there is a need for innovative optimization techniques to make such models scalable with respect to the population size, which presents another direction of research.

Structured inference for process parameters is yet another direction where more work needs to be done. Currently, the bands that are generated around the mean prediction curves (for the infected and deceased compartments) use some `common-sense' perturbations of the estimated population parameters (from training data), but clearly more objectivity is needed. As discussed previously, the bands need to be interpreted with a certain grain of salt. Appropriate calibration of the variability of the obtained estimates remains an interesting and open question. 

\subsection{Caveats}
We end our discussion with some caveats that are critical to keep in mind while interpreting the results of our research and similar other works on Covid predictions and analyses. While the model presented in this paper generally exhibits good fit to the training data and reasonable projections for the test period (to the extent that such data are available, since at the time of writing (in July) future data from August which constitutes one month of test data for the southern states are unavailable), some of the simplifying assumptions that were made in the interests of tractability should be mentioned. 

1. Proper modeling of active case counts depends heavily on who receives the tests and how the number of tests changes over time \cite{silver2020why}.  The testing mechanism in this paper is somewhat idealized: an individual is only tested once and further it is assumed that the compartment status of an individual is determined at the end of the day on which they are tested. This latter assumption ignores delays in getting test results which were somewhat of an issue especially in the earlier phases of the epidemic.  Also, we have made the assumption that tests give precise results. We have not taken into account the false positive and negative rates of tests in our model. 

2. In reality, patients and hospitals are not distributed uniformly across each state. Rather, there are usually centers of high concentration of infected cases, especially in crowded urban areas. Similarly, economic lockdowns do not always apply to whole states but rather strict lockdowns are enforced only on centers of spread at the county level. This, in turn, affects the cost and duration of lockdowns. In this work, we consider a state as one single unit, and therefore no differentiation is made at more granular levels (e.g. counties). However, as pointed out in the previous subsection, a spatio-temporal model for  a state where one treats counties as units of location, can alleviate some of these issues.

3. In our analysis the economic cost of lockdowns is assumed to increase linearly in the duration of lockdown. This  assumption can be violated in practice as longer lockdowns tend to have longer lasting and more devastating effects. For example, many businesses may be able to endure one week of lockdown, say with a cost $c$ to the economy. However, a lockdown of 4 weeks could make the business shutter permanently, leading to a cost much higher than $4c$ to the economy. Furthermore, the loss function that was optimized to obtain recommended policies, while designed to trade-off the natural losses against each other, can certainly be embellished with more input from economists and government policy makers. Our loss function is meant to serve as a broad template, but depending on the specific applications and goals, other fine-tuned loss functions can certainly be used. 

\begin{figure}[H] % "[t!]" placement specifier just for this example
\begin{subfigure}{0.48\textwidth}
\includegraphics[width=\linewidth]{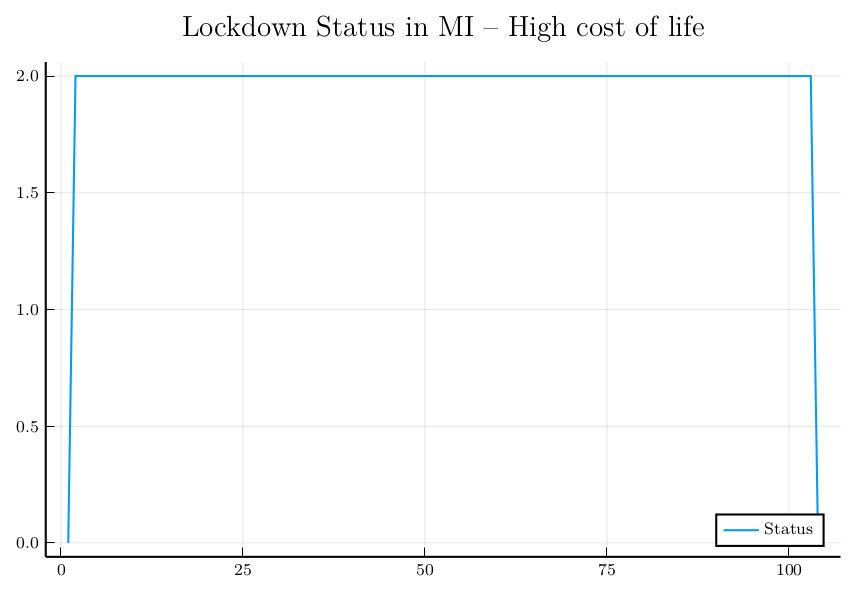}
\caption{} \label{fig:a}
\end{subfigure}\hspace*{\fill}
\begin{subfigure}{0.48\textwidth}
\includegraphics[width=\linewidth]{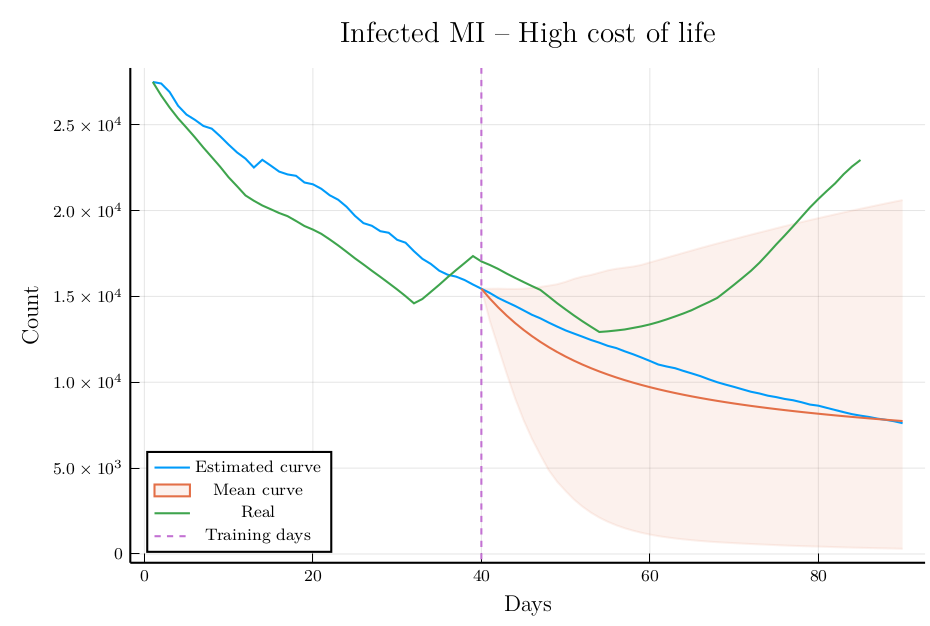}
\caption{} \label{fig:b}
\end{subfigure}

\medskip
\begin{subfigure}{0.48\textwidth}
\includegraphics[width=\linewidth]{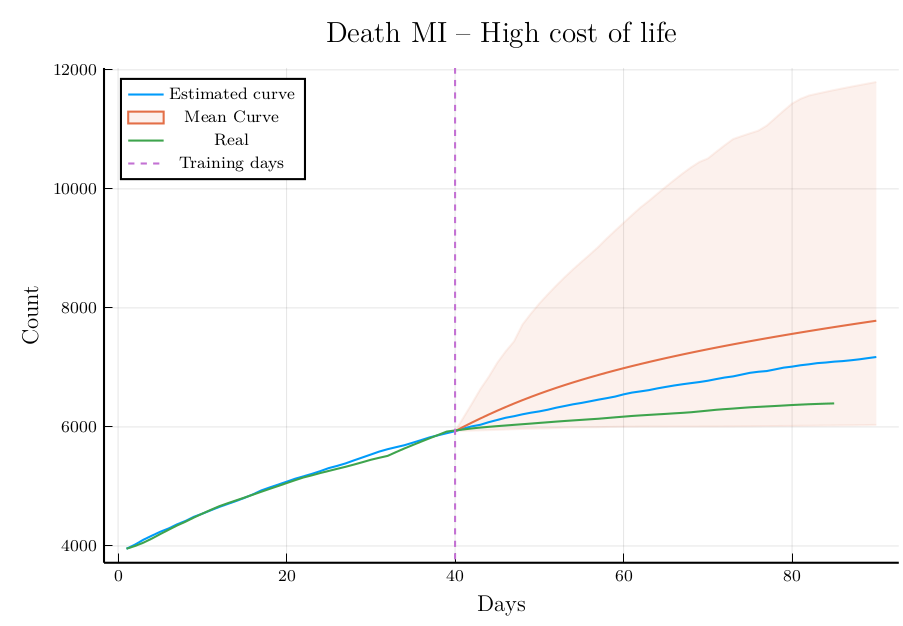}
\caption{} \label{fig:c}
\end{subfigure}\hspace*{\fill}
\begin{subfigure}{0.48\textwidth}
\includegraphics[width=\linewidth]{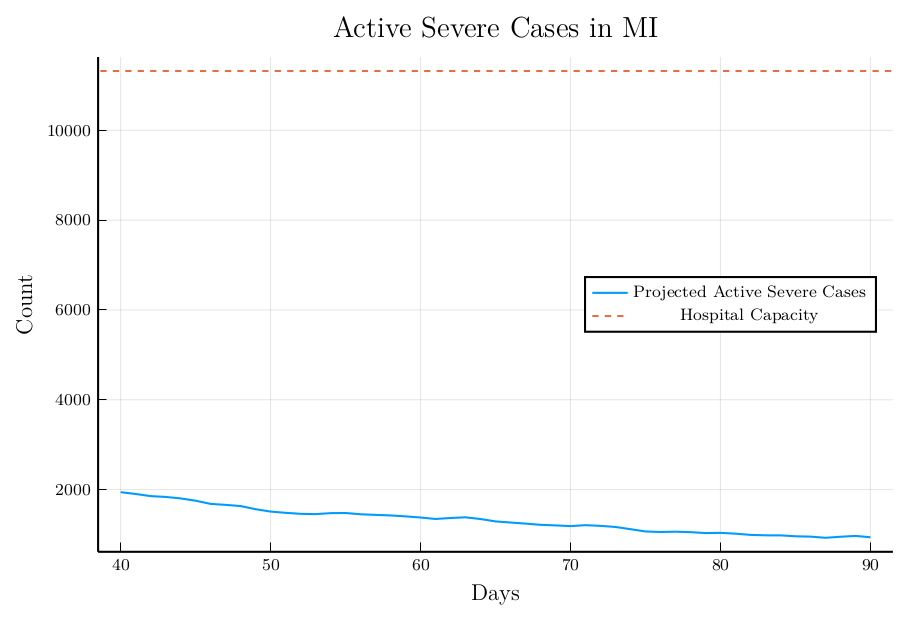}
\caption{}
\end{subfigure}\hspace*{\fill}
\caption{Perturbation analysis of MI with the proportion of severe cases $p_S = 0.15$.} \label{fig:perturb}
\end{figure}

4. As evidenced by the sensitivity bands in our plots, the projections of the model are sensitive to parameter specifications.  Consequently, the optimal lockdown policy can change sharply for slight changes in parameter estimates. As an example, we repeated our analysis for MI (for the high value of cost of life $4.7$) with $p_S = 0.15$\footnote{This corresponds to the higher hospitalization rates seen in April, while the May rates were lower.} and the other parameters set or optimized as before. In this perturbation scenario, we obtain full lockdowns (status = 2) for the whole 90-day period under consideration (see Figure \ref{fig:perturb}) whereas previously (with $p_S = 0.07$) the optimal policy enforced no lockdown beyond the training period for high and low cost of life. 
Thus, any policy specification should be examined and considered carefully by policy-makers, possibly with some sensitivity analysis, and also keeping in mind the situation on the ground which the modeler has no access to when they are making \emph{predictions}.

5. Building upon the previous point, an important question arises as to how far down the road predictions of such models should be relied upon. While we have predicted up to a period of 50 days, in practice we recommend updating predictions much before a 7 week duration, since prediction errors can increase quickly with longer time horizons and the consequence of mis-matched predictions (generally unavoidable) is often human lives. One may want to update predictions once every 2 weeks (especially if the spread is unchecked over a certain region), or 4 weeks (if the spread is generally well controlled). 
The dynamics of contagious disease propagation involve a slew of \emph{intangibles and imponderables} that no mathematical model can really capture, since human behavior is quite non-robust to shocks. For example, turmoil in the US over both racial issues and political partisanship has manifested itself in  massive demonstrations and public rallies as well as multiple civilian-police confrontations in recent months. Such mass-events have the potential to cause spikes in infection and change the subsequent dynamics of the disease. Obviously, no degree of quantitative modeling can account for the exact nature of such disturbances. 
%This is a drastic and qualitative difference in the optimal policy that must be acknowledged when recommending policies for real-world application.

\section{Acknowledgements}
This work was done under the auspices of a grant awarded by the  MIDAS Propelling Original Data Science (PODS) pilot funding program at University of Michigan.  

\section{Supplements}
\subsection{Additional mathematical details}
\label{sec:supplement} 
We derive below the form of the transition probability from compartment S to L. At the outset let us define a few quantities and articulate the underlying assumptions:
\newline
\newline
By $R_{a(t)}$ we denote the basic reproduction number under lockdown status $a(t)$: it is on an average the number of people to whom an infected person transmits the virus over the number of days that he/she is capable of infecting [which for our model is the average number of days $D$ that they stay in state $I_m$] under the assumption that the proportion of susceptibles is close to 1. We assume that the number of people $N$ that a random person comes into contact with on a daily basis follows Poisson($\mu_{a(t)}$) where $\mu_{a(t)}$ varies by lockdown status. A strict lockdown corresponds to a small value for this number. If $\beta$ is the probability of transmission of infection per contact between an individual in state S and one in state $I_m$, then:
\[ R_{a(t)} = \mu_{a(t)} \beta D \,.\] 
Now, 
\begin{align*}
P_t^{S \to L} & = \P\left(\text{A person in S is infected at time $t+1$}\right) \\
& = \sum_{n=1}^{\infty}\P\left(\text{Infection } \mid \text{Meet}\;n \;\text{people} \right) \P\left(\text{Meet}\;n\; \text{people}\right) \\
& = \sum_{n=1}^{\infty}\left[\sum_{j=0}^n\dbinom{n}{j} \left( \frac{X_t(I_m)}{N}\right)^j \left(1 -\frac{X_t(I_m)}{N}\right)^{n-j}\left(1 - (1 - \beta)^j\right)\right]\frac{e^{-\mu_{a(t)}}\mu^n_{a(t)}}{n!} \\
& = \sum_{n=1}^{\infty}\left[1 - \left(1-\beta \frac{X_t(I_m)}{N}\right)^n\right]\frac{e^{-\mu_{a(t)}}\mu^n_{a(t)}}{n!} \\
& = 1- \sum_{n=1}^{\infty}\left(1-\beta \frac{X_t(I_m)}{N}\right)^n\frac{e^{-\mu_{a(t)}}\mu^n_{a(t)}}{n!} \\
& = 1 - e^{-\mu_{a(t)}\beta \frac{X_t(I_m)}{N}}
\end{align*}
The above formula admits a ready approximation in terms of the quantity $R_{a(t)}$ which is described next: consider the situation where the status $a(t)$ remains unchanged over the period of time during which an infected person is in state $I_m$.  On an average, the person meets $\mu_{a(t)}$ people every day of which a proportion $X_t(s)/N$ are susceptible and each such contact spreads the infection with probability $\beta$. Thus, the average number of people infected on day $t$ is $\mu_{a(t)} X_t(S)/N \beta$. The average infected over the entire period can therefore be approximated by summing this quantity over the average number of days in the $I_m$ compartment, which is $D$. If we assume that the proportion $X_t(S)/N$ does not change dramatically over this period, we have $\mu_{a(t)} \times (X_t(S)/N) \times \beta \times D$ as the average infected over the entire period. On the other hand, the average infected over the entire period is also approximately $R_{a(t)} \times X_t(S)/N$: this follows from the fact that we are working under the assumption that $a(t)$ is fixed over the duration of the infecting phase and the rough constancy of the proportion of susceptibles over this phase, leading us to: 
$$D \times \mu_{a(t)} \times \frac{X_t(S)}{N} \times \beta  = R_{a(t)} \times \frac{X_t(S)}{N} \,.$$
Hence we have: 
$$\mu_{a(t)} \times \beta  = \frac{R_0^{a(t)}}{D} \,.$$
Putting this back in the main equation we get: 
$$P_t^{S \to L} = 1 - e^{- \frac{R_0^{a(t)}}{D} \frac{X_t(I_m)}{N}} = 1 - \exp\left(-R_0^{a(t)} (P^{I_m \to I_s} + P^{I_m \to R})\frac{X_t(I_m)}{N} \right)\,,$$
upon noting that the number of days $D$ spent in state $I_m$ is a geometric random variable with mean $1/(P^{I_m \to I_s} + P^{I_m \to R})$.

\subsection{Additional plots}
In this subsection we include the plots for FL, AZ and NJ. 
\begin{figure}[H] % "[t!]" placement specifier just for this example
\begin{subfigure}{0.48\textwidth}
\includegraphics[width=\linewidth]{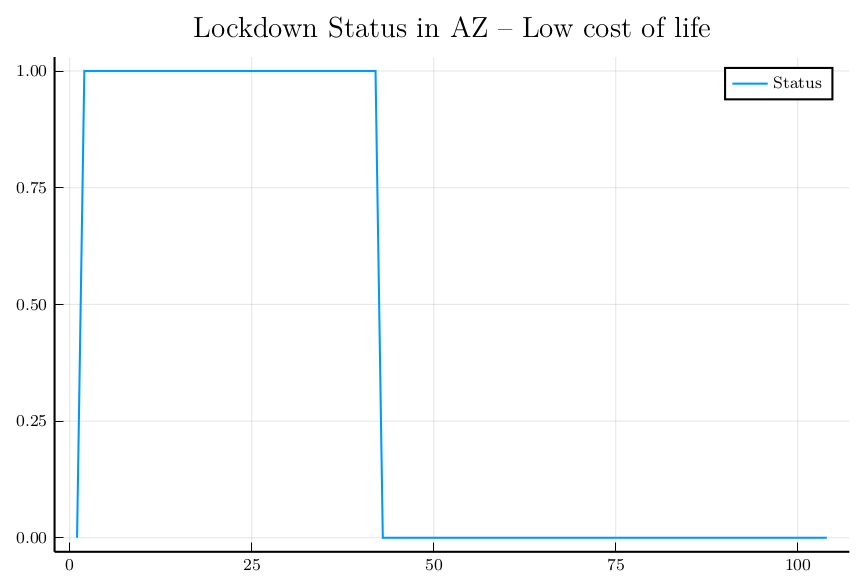}
\caption{} \label{fig:a}
\end{subfigure}\hspace*{\fill}
\begin{subfigure}{0.48\textwidth}
\includegraphics[width=\linewidth]{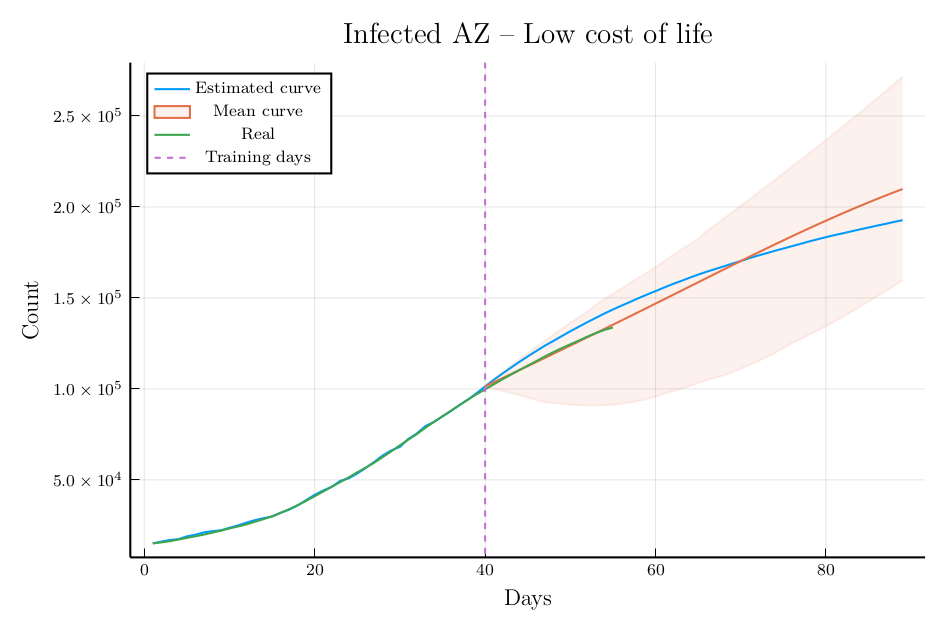}
\caption{} \label{fig:b}
\end{subfigure}

\medskip
\begin{subfigure}{0.48\textwidth}
\includegraphics[width=\linewidth]{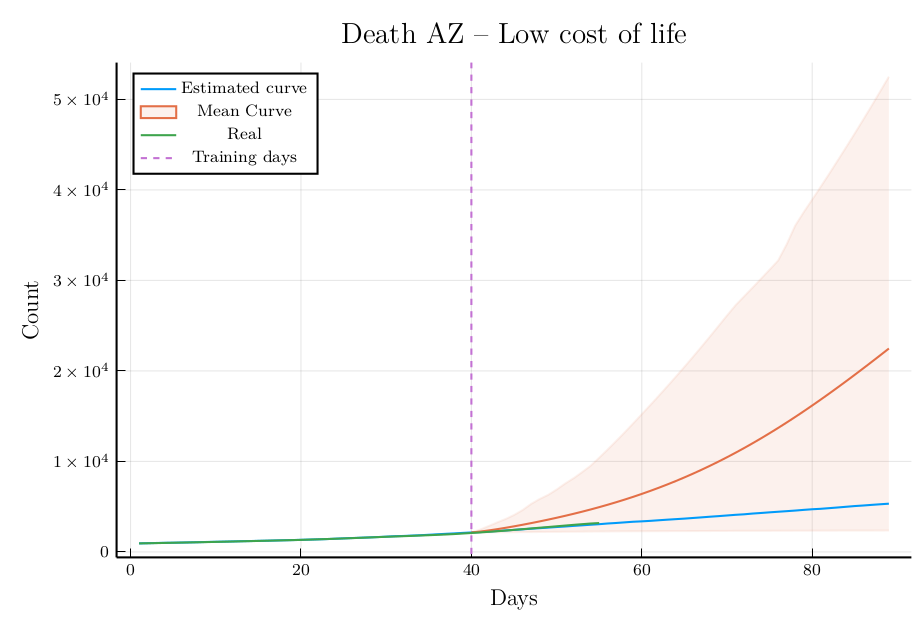}
\caption{} \label{fig:c}
\end{subfigure}\hspace*{\fill}
\begin{subfigure}{0.48\textwidth}
\includegraphics[width=\linewidth]{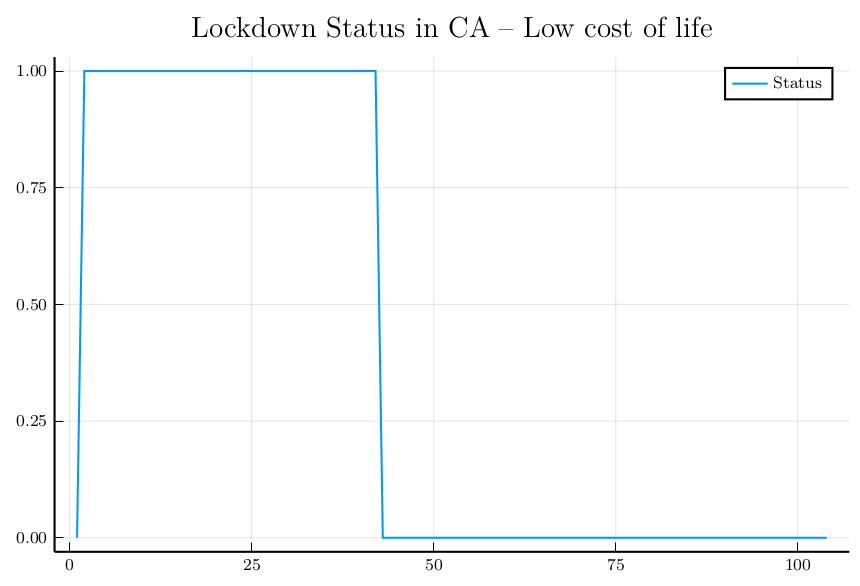}
\caption{}
\end{subfigure}\hspace*{\fill}

\begin{subfigure}{0.48\textwidth}
\includegraphics[width=\linewidth]{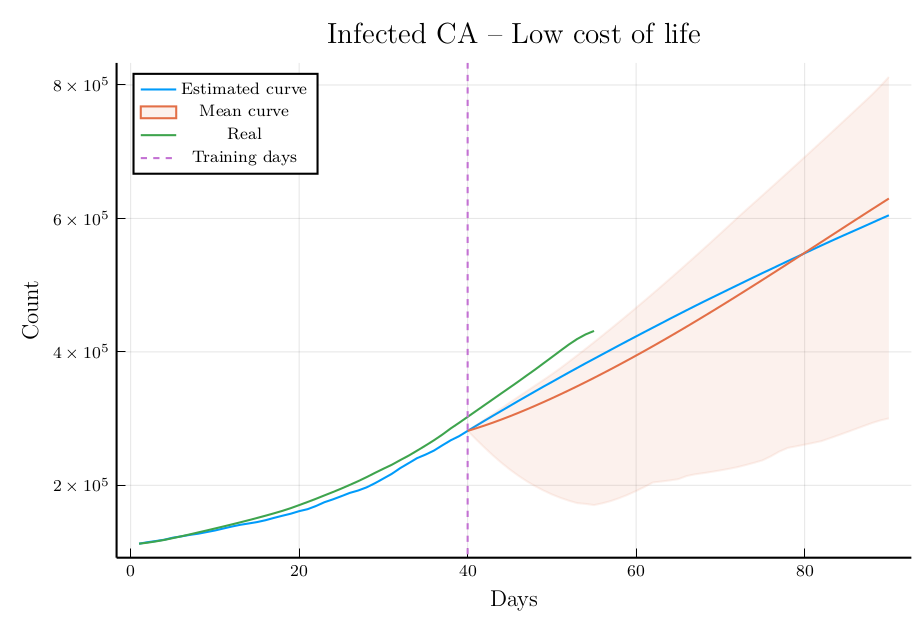}
\caption{}
\end{subfigure}
\medskip
\begin{subfigure}{0.48\textwidth}
\includegraphics[width=\linewidth]{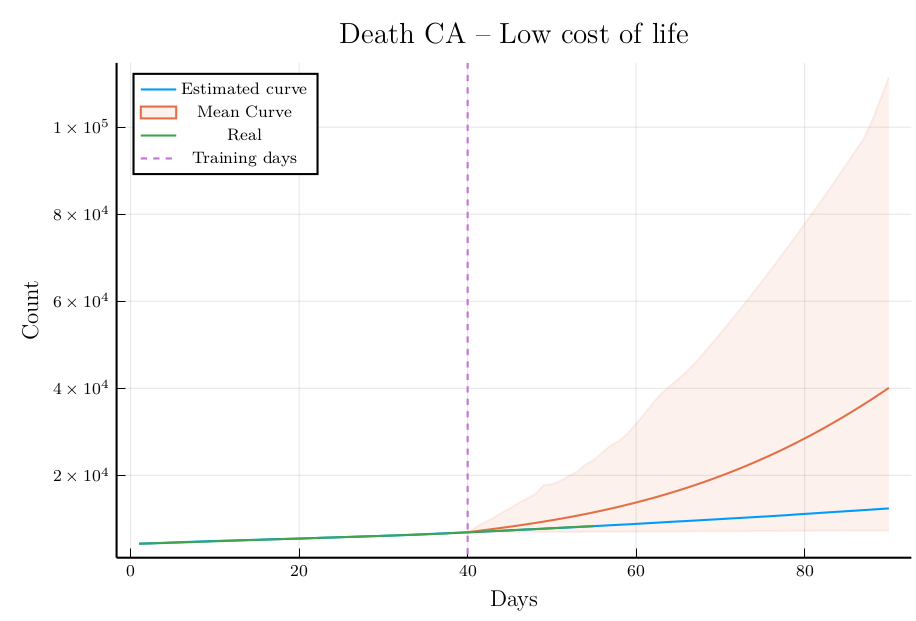}
\caption{}
\end{subfigure}

\caption{Analysis of AZ and CA} \label{fig:1}
\end{figure}

\begin{figure}

\begin{subfigure}{0.48\textwidth}
\includegraphics[width=\linewidth]{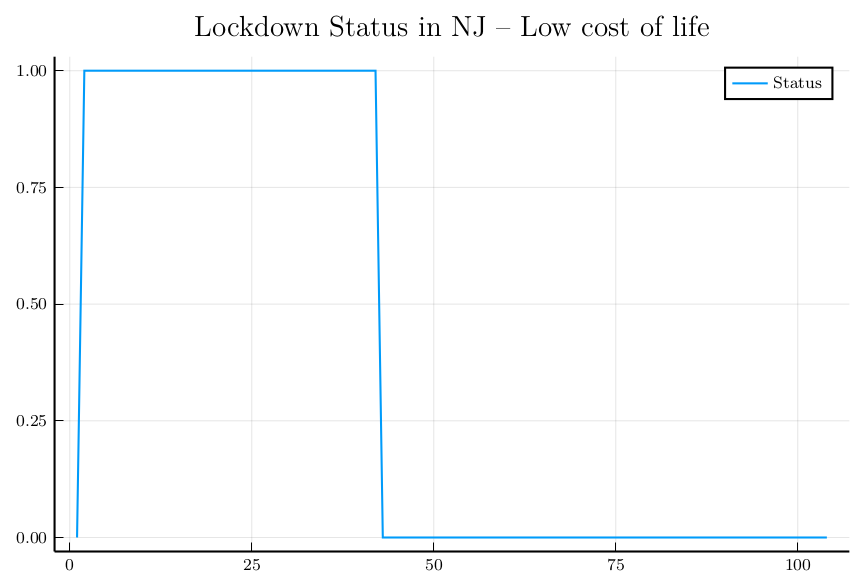}
\end{subfigure}\hspace*{\fill}
\begin{subfigure}{0.48\textwidth}
\includegraphics[width=\linewidth]{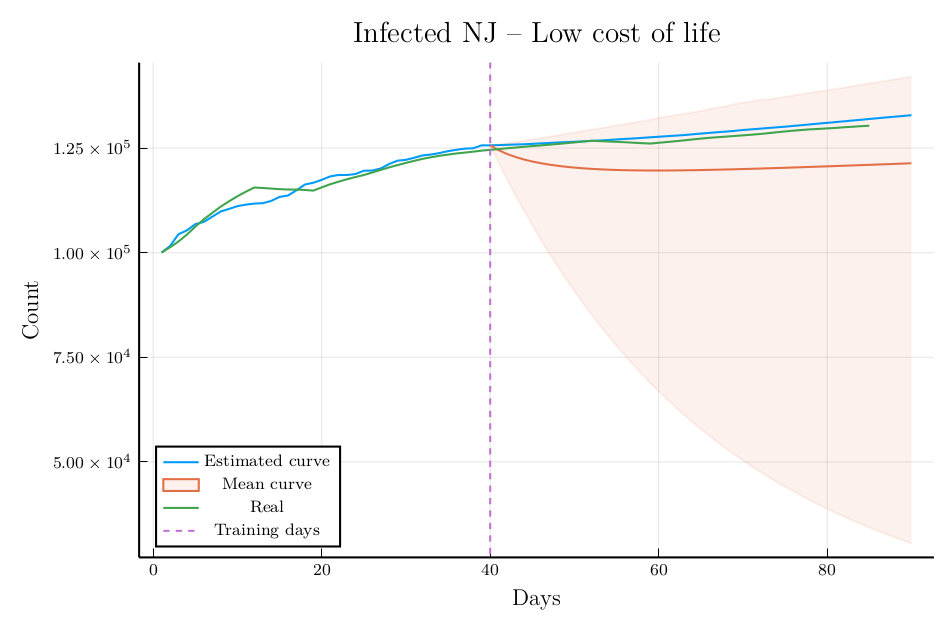}
\end{subfigure}
\end{figure}

\begin{figure}[H]
\centering 
\includegraphics[width=0.5\linewidth]{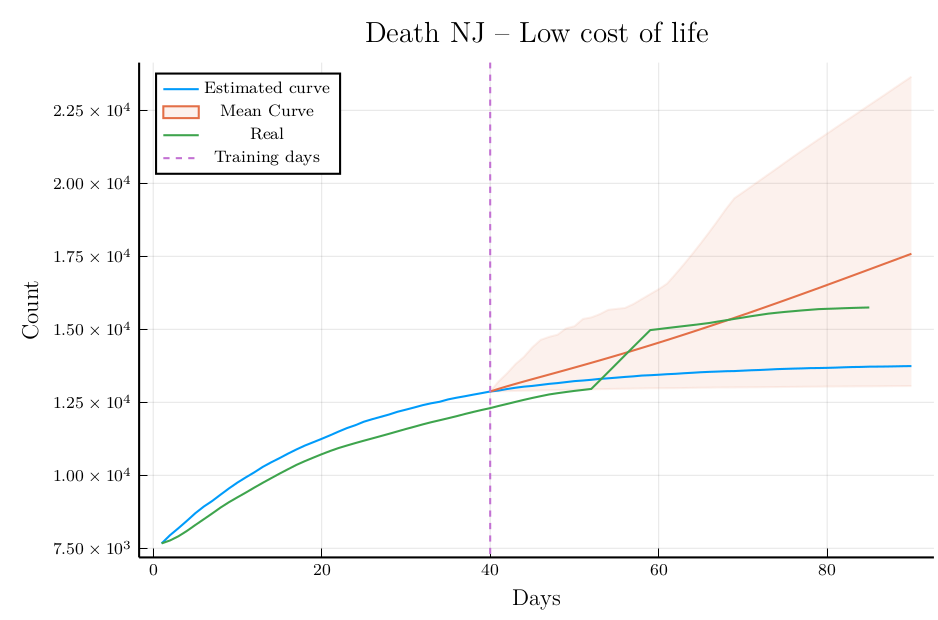}
\caption{Analysis of NJ}
\end{figure}

\begin{figure}
\begin{subfigure}{0.48\textwidth}
\includegraphics[width=\linewidth]{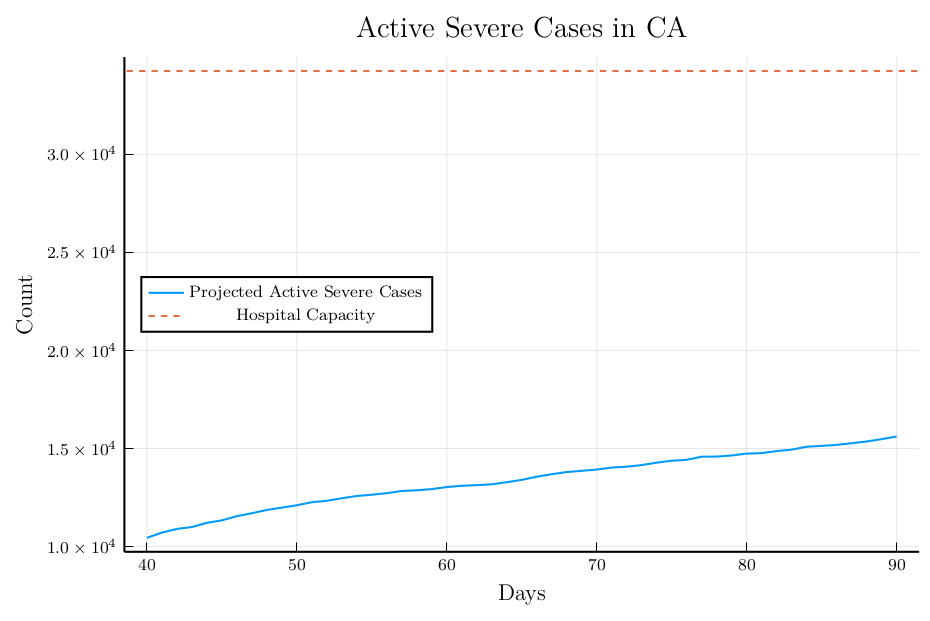}
\caption{} \label{fig:d}
\end{subfigure}
\medskip
\begin{subfigure}{0.48\textwidth}
\includegraphics[width=\linewidth]{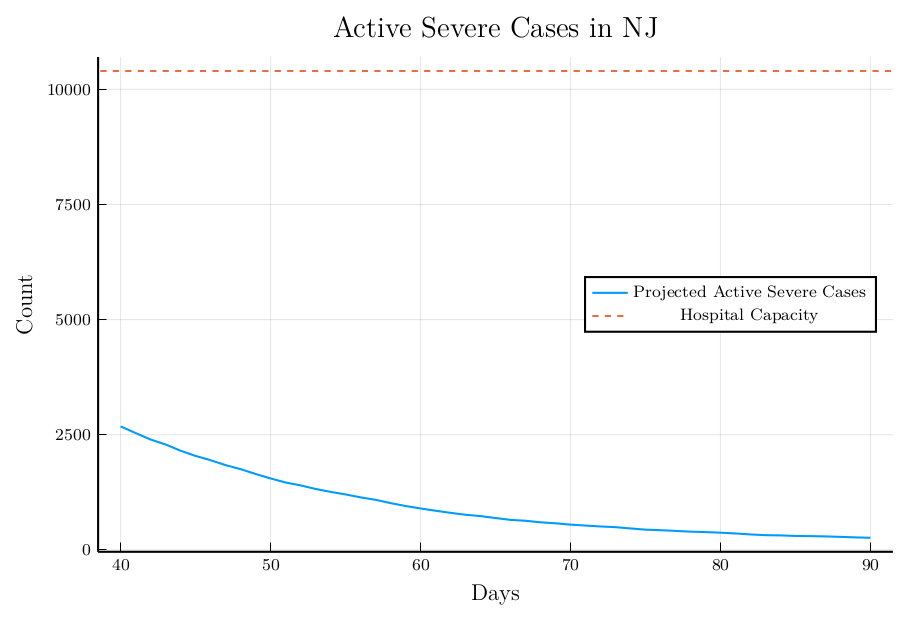}
\caption{} \label{fig:e}
\end{subfigure}\hspace*{\fill}

\begin{subfigure}{0.48\textwidth}
\includegraphics[width=\linewidth]{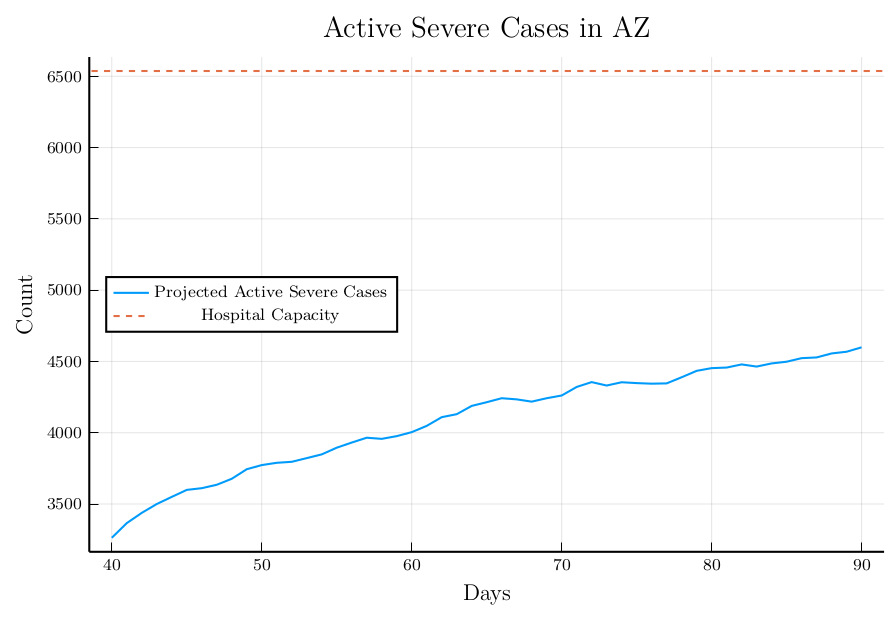}
\caption{} \label{fig:f}
\end{subfigure}

\caption{Severe Cases vs. Hospital Capacity}
\end{figure}

\bibliography{bibfile}{}
\bibliographystyle{plain}
\end{document}